\documentclass[aip,superscriptaddress]{revtex4}
\usepackage{graphicx}
\usepackage{fancyhdr}
\usepackage{amsmath} 
\usepackage{bm}
\usepackage{amsxtra}
\usepackage{amssymb}
\usepackage{amsthm}
\usepackage{latexsym}
\usepackage{lscape}
\RequirePackage{xspace}
\usepackage{epsfig,color,colordvi,amsfonts,multirow,rotating,relsize}
\usepackage{subfigure}
\usepackage{adjustbox}

\def\pt        {\mbox{$p_T$}\xspace}

\newcommand{\Pp}{\mathswitchr p}
\def\mathswitch#1{\relax\ifmmode#1\else$#1$\fi}
\def\mathswitchr#1{\relax\ifmmode{\mathrm{#1}}\else$\mathrm{#1}$\fi}
\def\mathswitchit#1{\relax\ifmmode{#1}\else$#1$\fi}
\newcommand{\TeV}{\unskip\,\mathrm{TeV}}
\newcommand{\GeV}{\unskip\,\mathrm{GeV}}
\newcommand{\PW}{\mathswitchr W}
\newcommand{\PZ}{\mathswitchr Z}

\newcommand{\MZ}{\mathswitch {M_\PZ}}

\newcommand{\pba}{\unskip\,\mathrm{pb}}

\newcommand{\born}{{\mathrm{Born}}}

\def\ga{\gamma}
\def\de{\delta}

\def\si{\sigma}

\newcommand{\EW}{{\mathrm{EW}}}
\newcommand{\QCD}{{\mathrm{QCD}}}

\newcommand{\veto}{{\mathrm{veto}}}

\newcommand{\full}{\mathrm{full}}
\newcommand{\rT}{{\mathrm{T}}}

\newcommand{\alphaw}{\ensuremath{\alpha_{\mathrm{w}}}}
\newcommand{\alphas}{\ensuremath{\alpha_{\mathrm{s}}}}

\newcommand{\ensuremathr}[1]{\ensuremath{\mathrm{#1}}}
\newcommand{\rw}{\ensuremathr{w}}

\newcommand{\order}[1]{\ensuremath{ {\mathcal{O}\left( #1 \right)} }}
\def\met {\ensuremath{{E\!\!\!/}_{\mathrm{T}}}\xspace}

\begin{document}

\title{Electroweak Corrections at High Energies}

%

\author{Kalanand Mishra}
\affiliation{Fermi National Accelerator Laboratory, Batavia, IL 60510, USA}
\author{Luca Barz\`e}
\affiliation{CERN, PH-TH Department, Geneva, Switzerland}
\author{Thomas Becher}
\affiliation{Albert Einstein Center for Fundamental Physics, Institut f\"ur Theoretische Physik, Universit\"at Bern, Sidlerstrasse 5, CH-3012 Bern, Switzerland}
\author{Mauro Chiesa}
\affiliation{Dipartimento di Fisica, Universit\`a di Pavia, and \\
INFN, Sezione di Pavia, Via A. Bassi 6, 27100 Pavia, Italy}
\author{Stefan Dittmaier}
\affiliation{Albert-Ludwigs-Universit\"at Freiburg, Physikalisches Institut, D-79104 Freiburg, Germany}
\author{Xavier Garcia i Tormo}
\affiliation{Albert Einstein Center for Fundamental Physics, Institut f\"ur Theoretische Physik, Universit\"at Bern, Sidlerstrasse 5, CH-3012 Bern, Switzerland} 
\author{Alexander Huss}
\affiliation{Albert-Ludwigs-Universit\"at Freiburg, Physikalisches Institut, D-79104 Freiburg, Germany} 
\author{Tobias Kasprzik}
\affiliation{Karlsruhe Institute of Technology, Institut f\"ur Theoretische Teilchenphysik, D-76128 Karlsruhe, Germany}
\author{Ye Li}
\affiliation{SLAC National Accelerator Laboratory, Stanford University, Stanford, CA 94309, USA}
\author{Guido Montagna}
\affiliation{Dipartimento di Fisica, Universit\`a di Pavia, and \\
INFN, Sezione di Pavia, Via A. Bassi 6, 27100 Pavia, Italy}
\author{Mauro Moretti}
\affiliation{Dipartimento di Fisica e Scienze della Terra, Universit\`a di Ferrara, and \\
INFN, Sezione di Ferrara, Via Saragat 1, 44100 Ferrara, Italy}
\author{Oreste Nicrosini}
\affiliation{Dipartimento di Fisica, Universit\`a di Pavia, and \\
INFN, Sezione di Pavia, Via A. Bassi 6, 27100 Pavia, Italy}
\author{Frank Petriello}
\affiliation{High Energy Physics Division, Argonne National Laboratory, Argonne, IL 60439, USA} 
\affiliation{Department of Physics \& Astronomy, Northwestern University, Evanston, IL 60208, USA}
\author{Fulvio Piccinini}
\affiliation{Dipartimento di Fisica, Universit\`a di Pavia, and \\
INFN, Sezione di Pavia, Via A. Bassi 6, 27100 Pavia, Italy}
\author{Francesco Tramontano}
\affiliation{Dipartimento di Scienze Fisiche, Universit\`a di Napoli ``Federico II", and \\
INFN, Sezione di Napoli, via Cintia, 80126 Napoli, Italy
}

\begin{abstract}
We present a survey of the most abundant processes at the LHC 
for sensitivity to electroweak corrections at $\sqrt{s} = 8, 14, 33,$  
and 100~TeV proton-proton collision energies. The processes studied 
are pp$\to$ dijet, inclusive W and Z, W/Z+jets, and WW. 
In each case we compare the experimental uncertainty in the 
highest kinematic regions of interest with the relative magnitude of  
electroweak corrections and fixed-order $\alpha_S$ calculations. 
\end{abstract}

\maketitle

\thispagestyle{fancy}
 

\section{Introduction}
At the Large Hadron Collider (LHC) design energy of $14~\TeV$ 
and at future proton-proton (pp) colliders operating at even higher energies, 
the kinematic reach for the most abundant processes like dijet, 
inclusive W and Z, W/Z + jets, and vector-boson pair production 
will go deeper into the 
TeV regime and will become sensitive to electroweak (EW) corrections. 
In spite of their suppression
by the small value of the coupling constant
$\alpha$, the EW corrections can become large in the
high-energy domain due to the appearance of Sudakov 
logarithms that result from the virtual exchange of soft or collinear
massive weak gauge bosons~\cite{sudakov:1956,Fadin:1999bq}. 
The leading term is given by
$\alphaw\ln^2\left(Q^2/M_\PW^2\right)$, 
where $Q$ denotes the energy scale of the hard-scattering process,
$M_\PW$ is the W-boson mass, and 
$\alphaw=\alpha/\sin^2\theta_\rw=e^2/(4\pi\sin^2\theta_\rw)$ 
with $\theta_\rw$ denoting the weak mixing angle.
In the case of massless gauge bosons, these
logarithms are canceled by the corresponding
real-emission corrections.
However, the masses of the W and Z gauge bosons provide a physical cut-off.
Therefore,  
radiation of real $\PW$ or $\PZ$ bosons can be experimentally reconstructed 
for the most part. Only a small fraction that remains 
unresolved can compensate for the virtual corrections~\cite{Baur2007}.
Thus, at high scales $\lvert Q^2\rvert\gg M_\PW^2$, which are accessible
at the LHC and future colliders, the Sudakov logarithms can produce large
negative corrections. 
Although negligible for integrated cross sections, these
corrections can reach $10{-}20\%$ for transverse momentum (\pt) or 
invariant mass values in the TeV range. 
In this report, we study the evolution of EW corrections 
from the LHC to future high energy pp colliders 
for two benchmark values of collider energy, namely, 33~TeV and 100~TeV.

\section{Dijet production}
Inclusive production of two or more jets at the LHC allows for a 
detailed study of QCD at TeV energies. It is also the main 
background for searches of new heavy particles from beyond the  
Standard Model (SM) physics decaying into jets. 
\begin{figure}[!htb]
\includegraphics[width=0.5\textwidth]{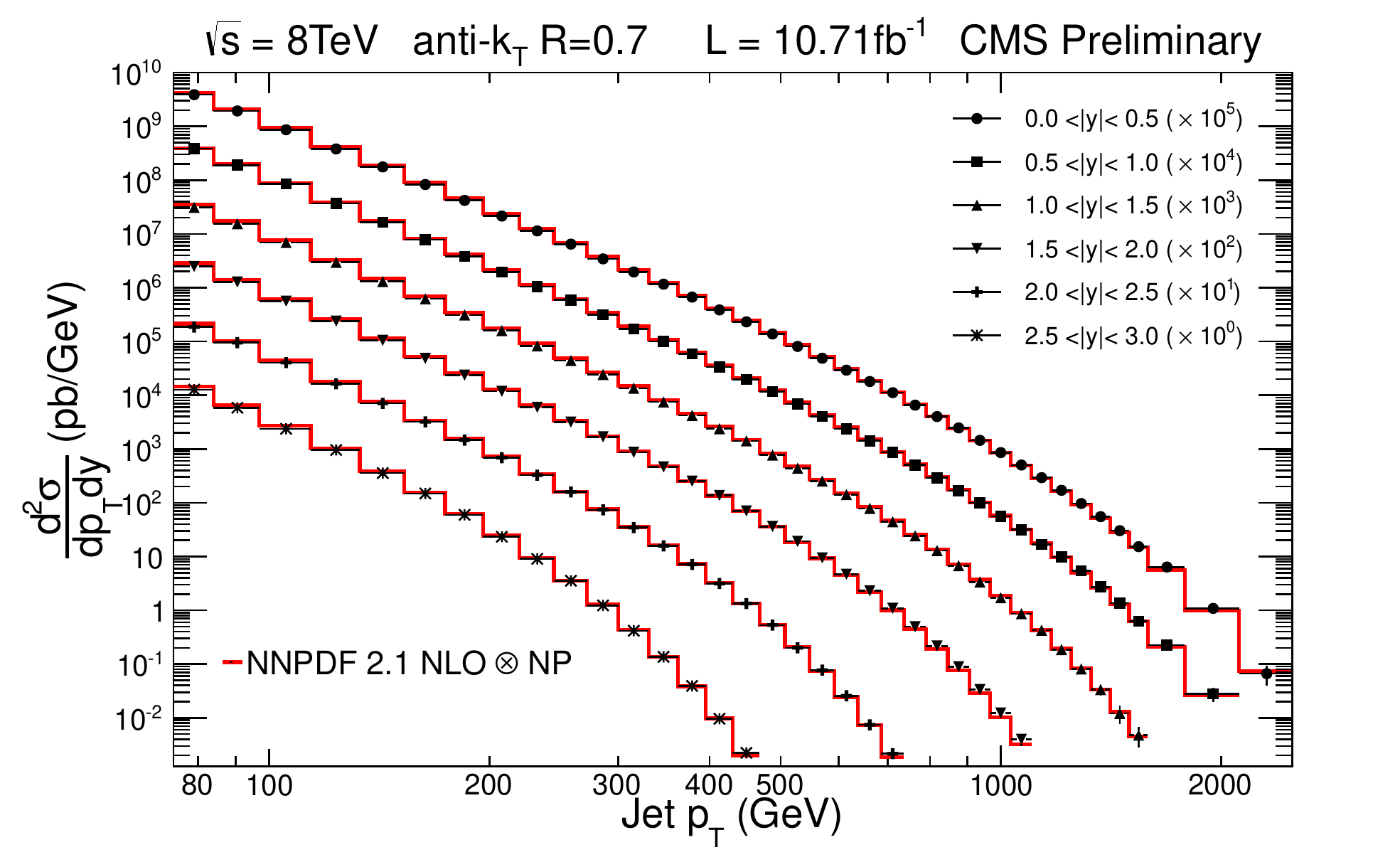}
\includegraphics[width=0.33\textwidth]{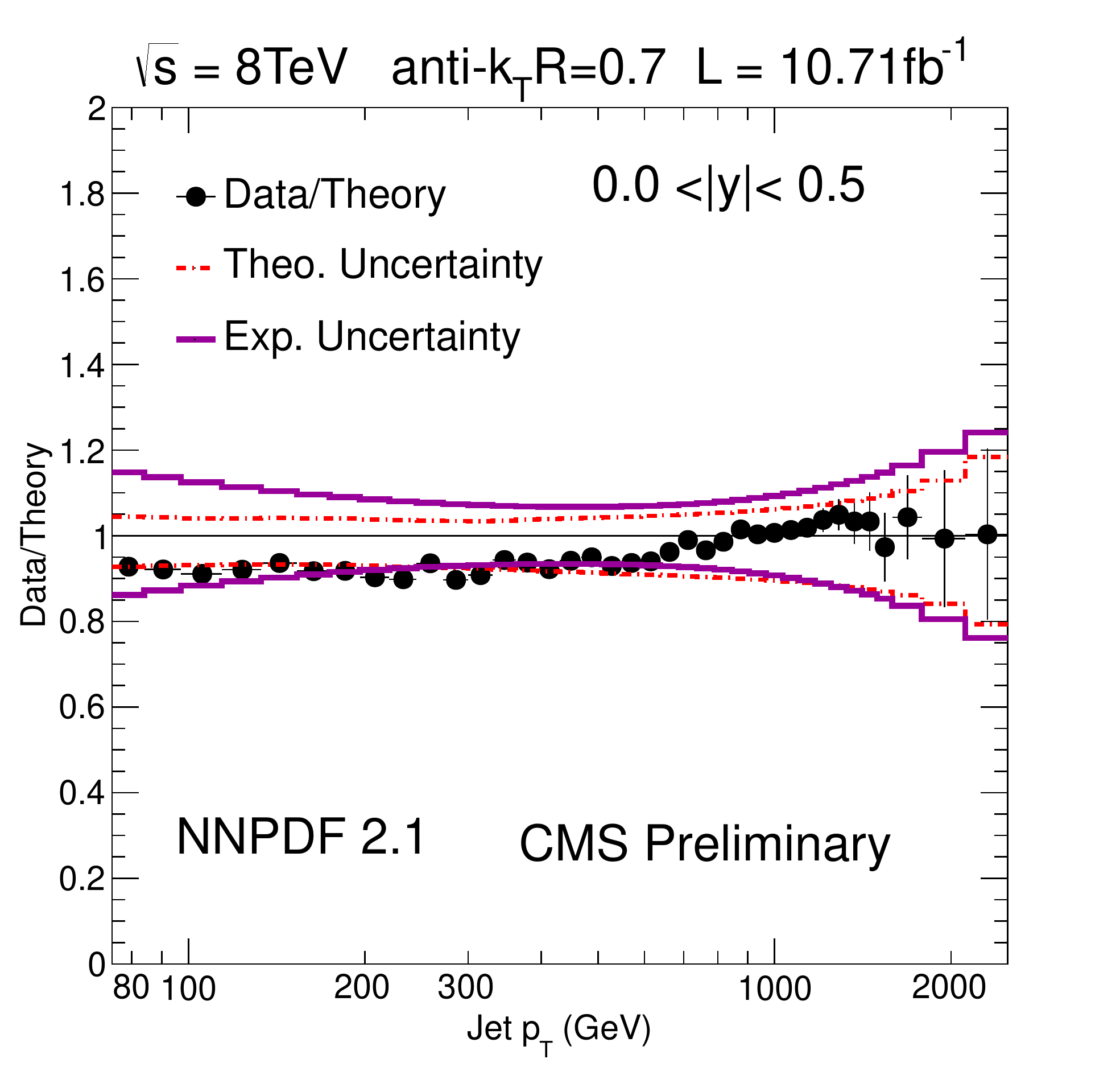}
\adjustbox{trim={.5\width} {.44\height} {0.05\width} {.33\height},clip}%
{\includegraphics[width=1\textwidth]{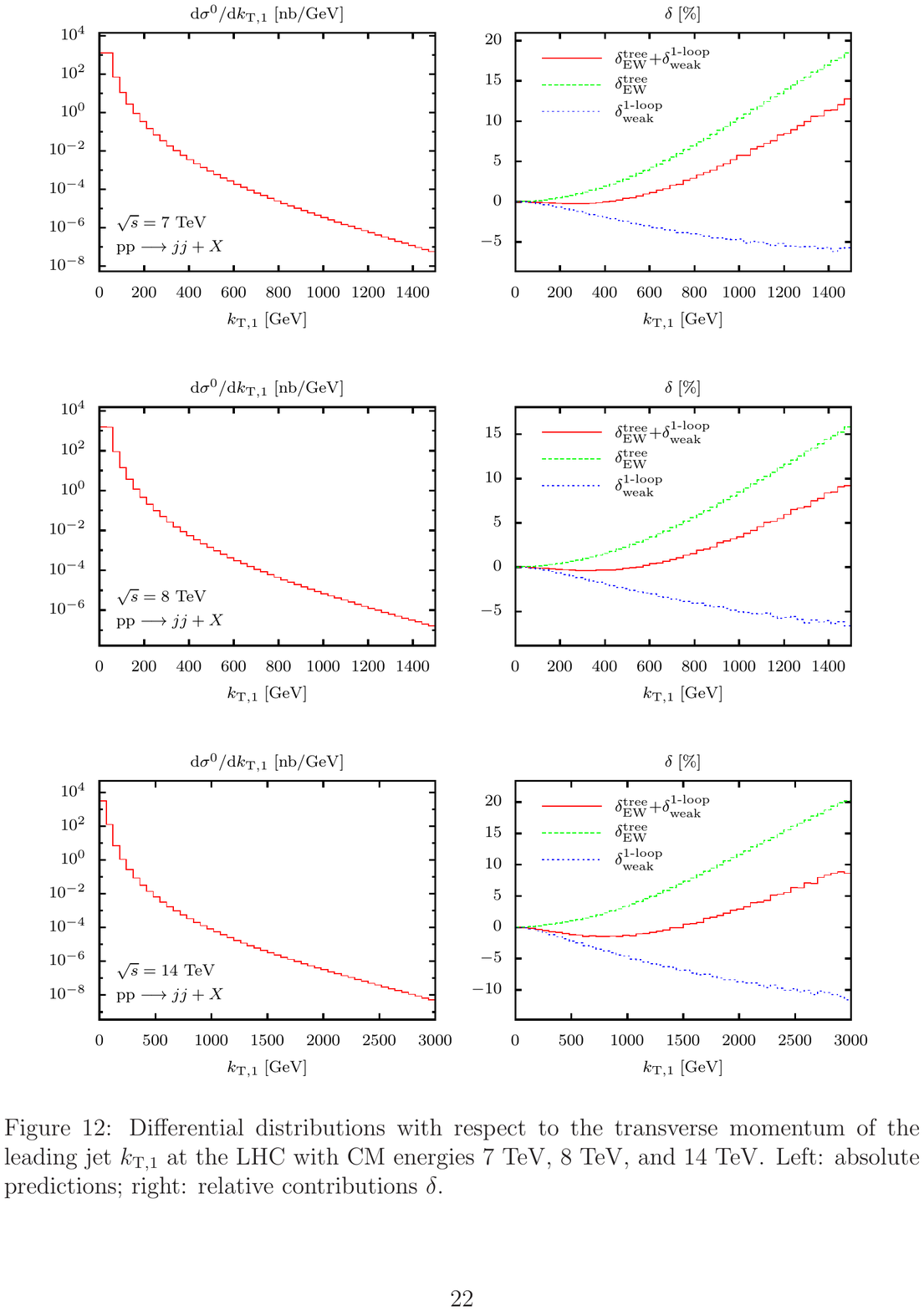}} 
\caption{The jet \pt spectrum measured in 8~TeV pp collisions (top left) along with uncertainties (top right) from Ref.~\cite{cms:dijet}. Also shown is the relative magnitude of EW corrections at $\sqrt{s} = 8~\TeV$ (bottom) taken from Ref.~\cite{Dittmaier:2012kx}.}
\label{fig:dijet_exp}
\end{figure}
Inclusive production of jets and dijets has been analyzed by the
ATLAS~\cite{Aad:2011fc} and CMS~\cite{cms:dijet} 
collaborations at $\sqrt{s} = 7$ and $8~\TeV$. 
These measurements show sensitivity to dijet invariant masses of up
to $5~\TeV$ and jet transverse momenta of up to $2~\TeV$
at the LHC. As shown in Fig.~\ref{fig:dijet_exp}, at the 
current level of experimental and theoretical accuracy,
the SM is able to describe data well. 
The size of the EW corrections is comparable 
to the experimental uncertainty for the highest \pt bins. 
In this respect the dijet measurements at $\sqrt{s} = 8~\TeV$ have already 
started probing the ``Sudakov zone''.  
\begin{figure}[!htb]
\begin{center}
\adjustbox{trim={.15\width} {.19\height} {0.1\width} {.15\height},clip}%
{\includegraphics[page=1,width=1\textwidth]{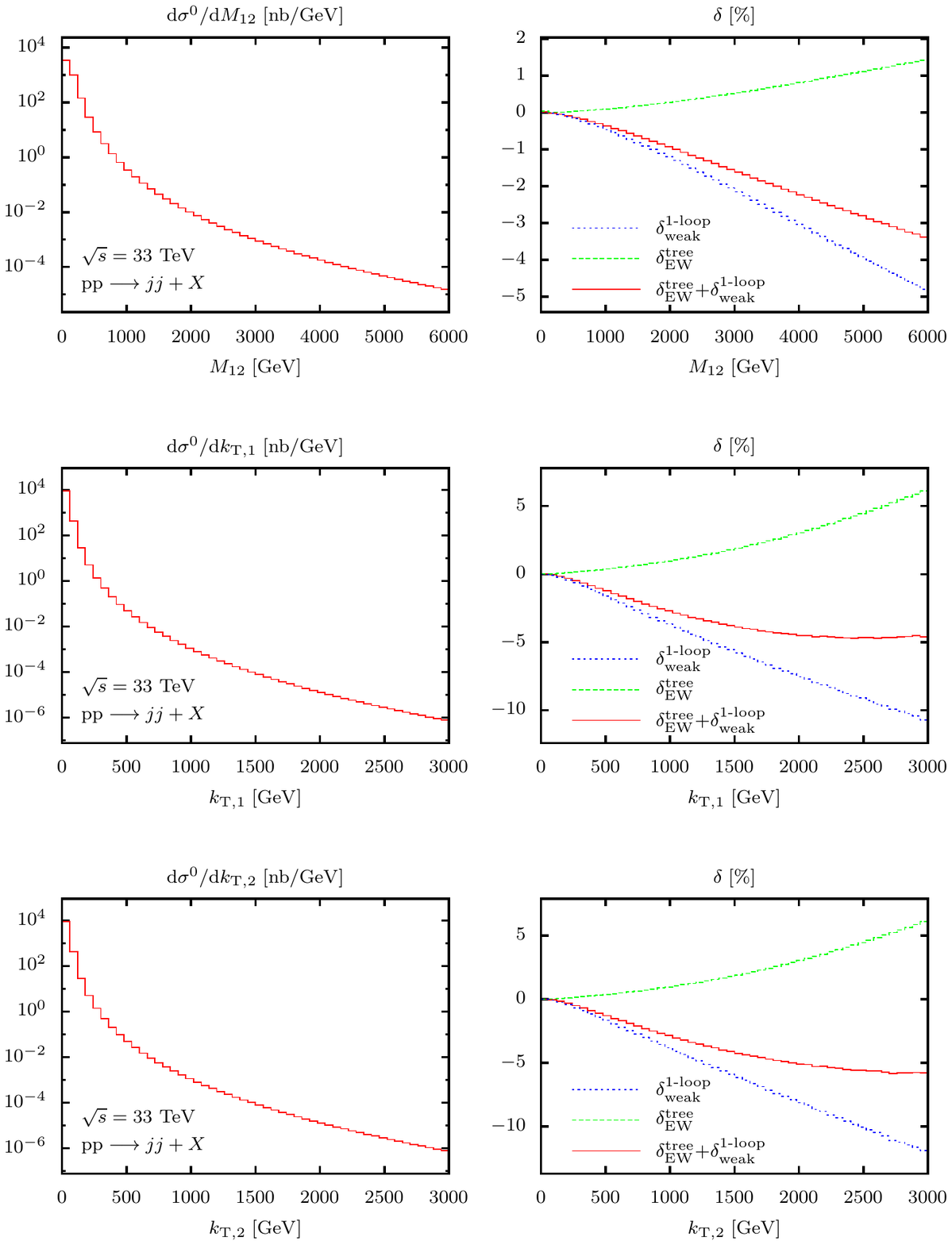}}
\caption{Dijet production: Differential distributions with respect to the 
    dijet invariant mass $M_{12}$~(top), the transverse momentum of the
    leading jet $k_{\text{T},1}$~(middle) and the 
    sub-leading jet $k_{\text{T},2}$~(bottom) at the
    pp collision energy of 33~TeV.
    Left: absolute predictions; right: relative contributions $\delta$.}
\label{fig:dijet_corr_33tev}
\end{center}
\end{figure}
\begin{figure}[!htb]
\begin{center}
\adjustbox{trim={.15\width} {.19\height} {0.1\width} {.15\height},clip}%
{\includegraphics[page=3,width=1\textwidth]{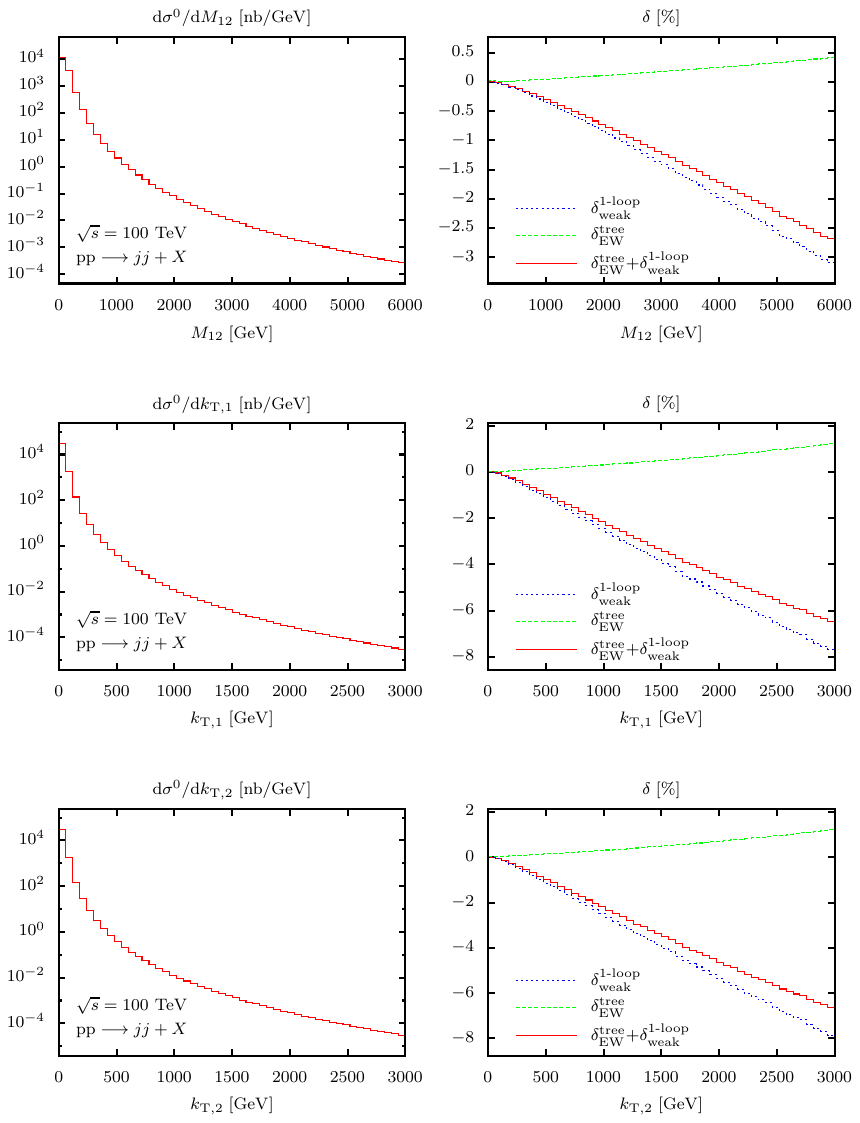}} 
\caption{Dijet production: Same distributions as in Fig.~\ref{fig:dijet_corr_33tev} for pp collision energy of 100~TeV.}
\label{fig:dijet_corr_100tev}
\end{center}
\end{figure}
The EW corrections to dijet production shown in 
Fig.~\ref{fig:dijet_exp}b are taken from Ref.~\cite{Dittmaier:2012kx}.
These corrections comprise electroweak contributions of
$\order{\alphas\alpha,\,\alpha^2}$ to the leading-order (LO) QCD prediction
as well as next-to-LO (NLO) corrections through the order $\alphas^2\alpha$. 
The tree-level contributions are of the same 
generic size as the loop corrections at $\sqrt{s} = 8~\TeV$. 
The total correction to the integrated cross section 
is negligible, typically staying below the per-cent level. 
However, the Sudakov logarithms 
affect the tails of the distributions in the dijet
invariant mass and in the transverse momenta of the two jets.
The magnitude of the corrections at  $\sqrt{s} = 8~\TeV$ and 14~TeV 
were found to be similar in Ref.~\cite{Dittmaier:2012kx}. 
Results from repeating the calculation for  
$\sqrt{s} = 33~\TeV$ and 100~TeV are shown in Fig.~\ref{fig:dijet_corr_33tev} 
and Fig.~\ref{fig:dijet_corr_100tev}, respectively.
We find that the 1-loop virtual corrections are not so 
dependent on the collider energy, while 
the tree-level corrections decrease with $\sqrt{s}$. 
Their cancellation is less perfect, and as a result, the 
virtual negative corrections dominate in the kinematic tails 
at large $\sqrt{s}$. Since the kinematic reach will increase 
with $\sqrt{s}$, these corrections will become progressively  
more important.

\section{Inclusive vector boson production}
\begin{figure}[!htb]
\begin{center}
\includegraphics[width=0.42\textwidth]{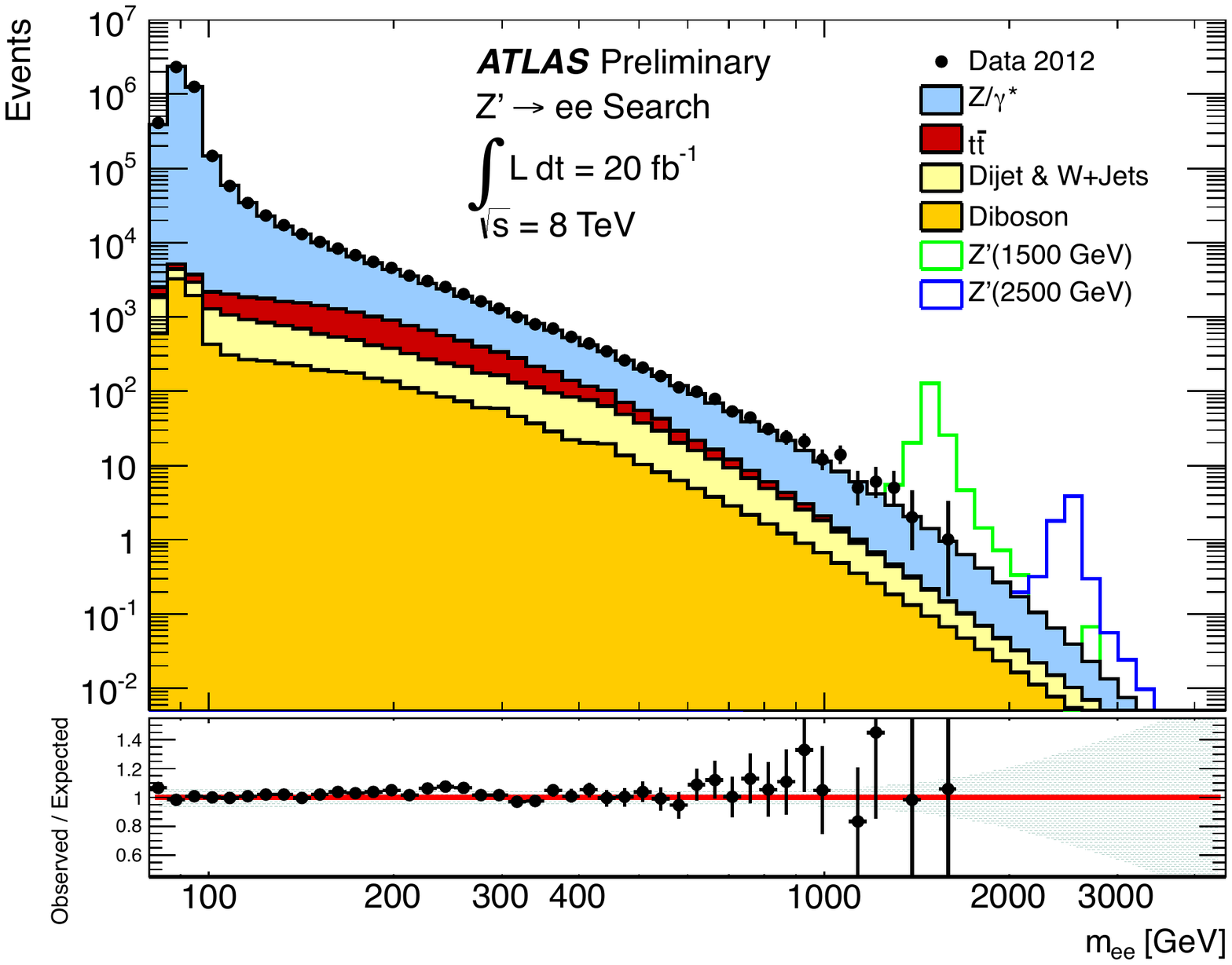}
\includegraphics[width=0.32\textwidth]{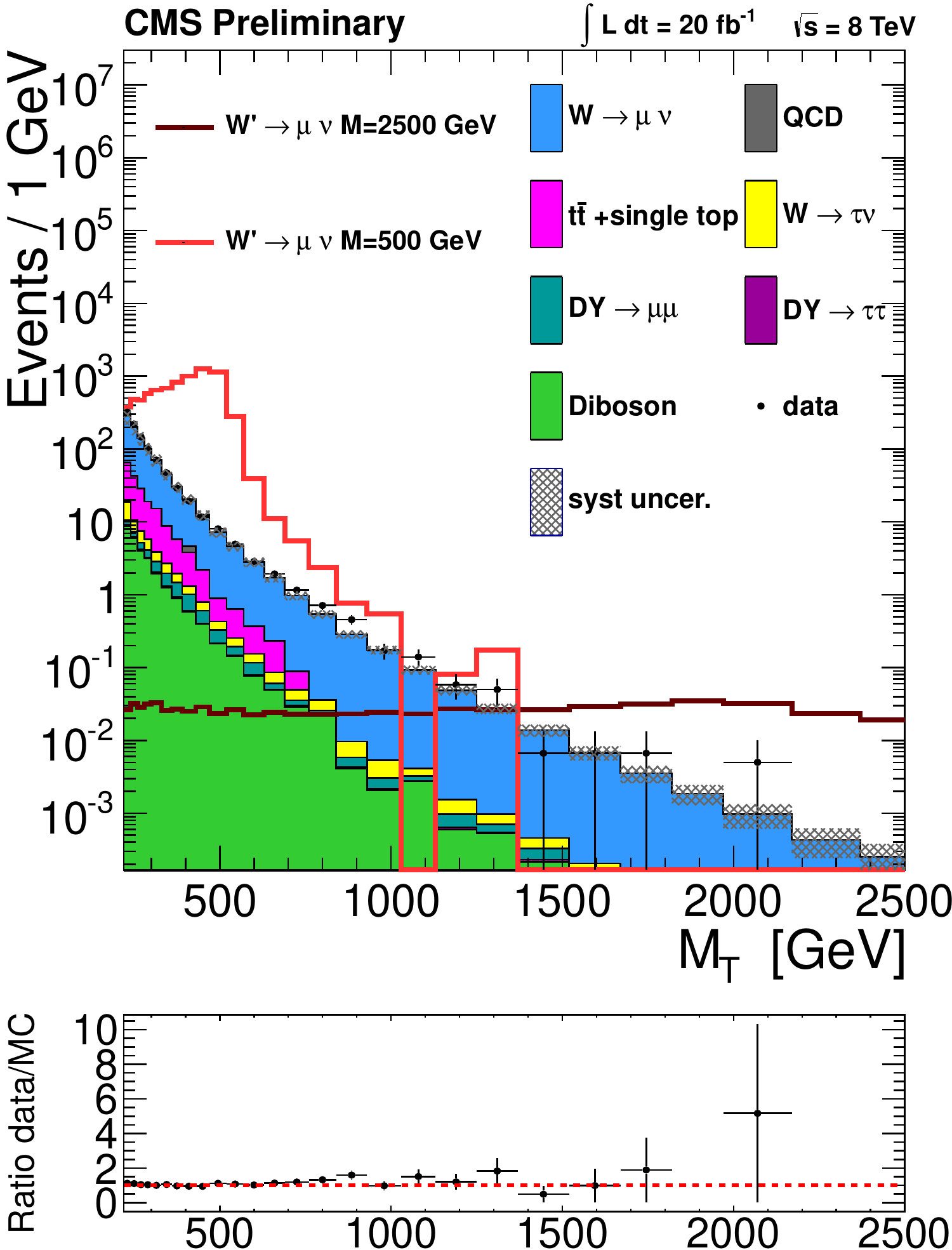}
\caption{Inclusive vector boson production: 
Observed vector-boson invariant mass distributions in 
8~TeV pp collisions compared to the SM expectation  
(left) in events containing two energetic electrons. 
  See Ref.~\cite{atlas:Zll} for details. 
  On the right, the transverse mass distribution of the muon + missing 
  transverse energy (\met) system  is shown for 
  events containing a high \pt muon and large \met. 
  See Ref.~\cite{cms:Wlnu} for details. 
}
\label{fig:V_exp}
\end{center}
\end{figure}
\begin{figure}[!htb]
\includegraphics[width=0.4\textwidth]{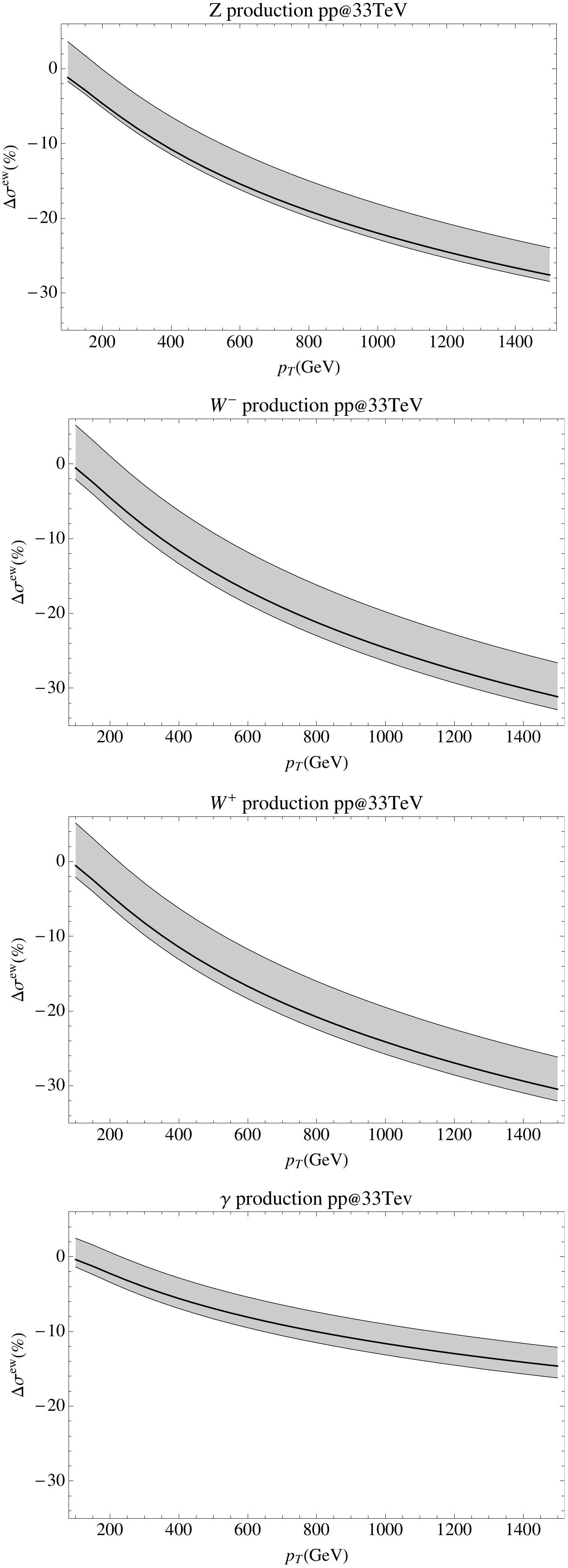}
\includegraphics[width=0.4\textwidth]{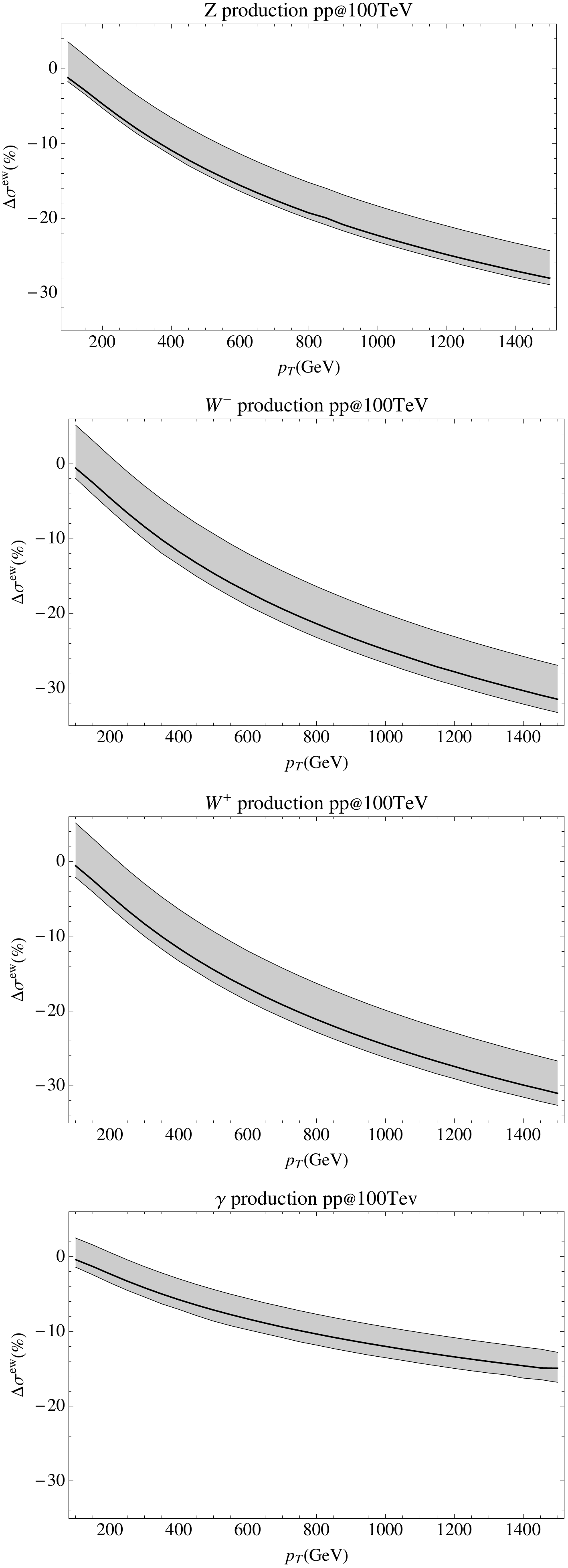}
\caption{Inclusive vector boson production: Relative contribution 
  from EW corrections as a function of  
  the vector boson \pt for Z~(row 1), $W^-$~(row 2), 
  $W^+$~(row 3), and $\gamma$~(row 4).
    Left: at the pp collision energy of 33~TeV; right: at 100~TeV.}
\label{fig:W_corr_100tev}
\end{figure}
The production of a single electroweak boson is one of the basic 
hard-scattering processes at the LHC and constitutes a major 
background to searches for physics beyond the SM, like W'$\to\ell\nu$ 
and Z'$\to\ell\ell$,  
and important SM measurements like H$\to$ZZ${}^*\to 4\ell$.
As in the dijet case, the EW corrections can become quite significant. 
Since we are considering single electroweak-boson
production, without additional radiation of soft or collinear
W or Z bosons, the cross section will contain 
Sudakov logarithms that can be as large as 20\% for boson 
$\pt\sim 1~\TeV$. 
These effects need to be included for precise predictions  
of kinematic distributions in the region $\pt\gg M_\PW$. 
Inclusive W and Z spectra have been analyzed by the
ATLAS~\cite{atlas:Zll} and CMS~\cite{cms:Wlnu} 
collaborations at $\sqrt{s} = 8~\TeV$, showing 
sensitivity to boson invariant masses of up
to $2~\TeV$ and \pt of up to $800~\GeV$.
As shown in Fig.~\ref{fig:V_exp}, at the 
current level of experimental and theoretical accuracy,
the SM is able to describe data well. 
The current experimental uncertainty for invariant masses above 
1~TeV is somewhat larger than the size of the EW corrections.  
The same measurements at $14~\TeV$ will be sensitive to 
probing the Sudakov zone.

For the above comparison we use Ref.~\cite{Becher:2013zua} which 
provides a computation of EW Sudakov effects in single W, Z, and
$\gamma$ production at large boson \pt at $\sqrt{s} = 7~\TeV$,  
with both QCD and electroweak effects included.
Switching off the QCD effects, and expanding to NNLO, the results of 
Ref.~\cite{Becher:2013zua} reproduce the earlier 
results ~\cite{Kuhn:2005az,Kuhn:2007cv}.
Results from repeating the same calculation for  
$\sqrt{s} = 33~\TeV$ and $100~\TeV$ are shown in Fig.~\ref{fig:W_corr_100tev}.
We find that the relative corrections are basically independent 
of collider energy and depend only on \pt. 
However, as the kinematic reach increases  
with $\sqrt{s}$, these corrections will become more important.

\subsection{Interplay of electroweak and QCD corrections in Drell-Yan production}

\begin{figure}[h!]
\begin{center}
\includegraphics[width=0.6\textwidth]{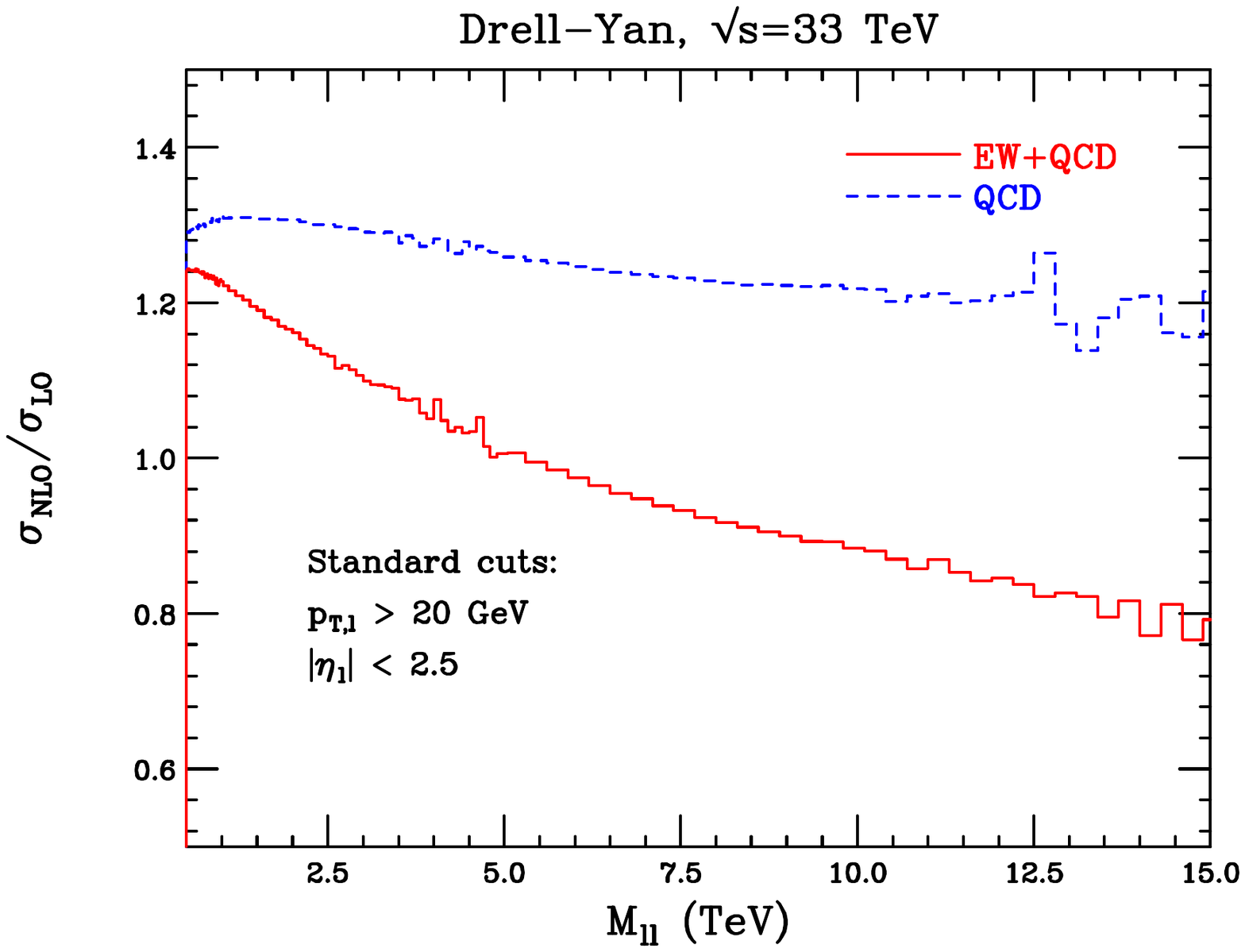}
\vspace{-1.5in}
\caption{Inclusive vector boson production: QCD corrections and combined electroweak-QCD corrections to lepton-pair production as a function of the lepton-pair invariant mass, at a 33~TeV $pp$ collider.}
\label{fig:qcd-EW-Fmll}
\end{center}
\end{figure}

The effects of electroweak Sudakov logarithms on more differential quantities, and their interplay with higher-order QCD corrections, are studied next for the example case of lepton-pair production via the Drell-Yan mechanism at a 33~TeV $pp$ collider.  The results shown are obtained with the numerical program FEWZ~\cite{Melnikov:2006kv,Gavin:2010az,Li:2012wna}, which additively combines higher-order QCD and electroweak corrections.  MSTW parton distribution functions~\cite{Martin:2009iq} at the appropriate order in QCD perturbation theory are used.  Shown first in Fig.~\ref{fig:qcd-EW-Fmll} is the lepton-pair invariant mass distribution, with minimal acceptance cuts imposed on the transverse momenta and pseudorapidities of the leptons.  The shift due to NLO QCD corrections alone is shown, as is the result of combining the full NLO electroweak correction with the QCD one.  Both shifts are normalized to the leading-order prediction.
Over a broad range of invariant masses, the QCD corrections increase the cross section by $20-30\%$.  The electroweak corrections grow in importance with invariant mass, and lead to a decrease in the cross section.  The electroweak corrections begin to overtake the QCD ones at $M_{ll} \approx 5~\TeV$.  The reduction in the cross section induced by the combined corrections reaches 20\% at invariant masses of 15~TeV.  Measurements of Drell-Yan production in 33~TeV collisions will be sensitive to the Sudakov zone over a large fraction of the available invariant mass range.

\begin{figure}[h!]
   \centering
   \includegraphics[width=0.49\textwidth]{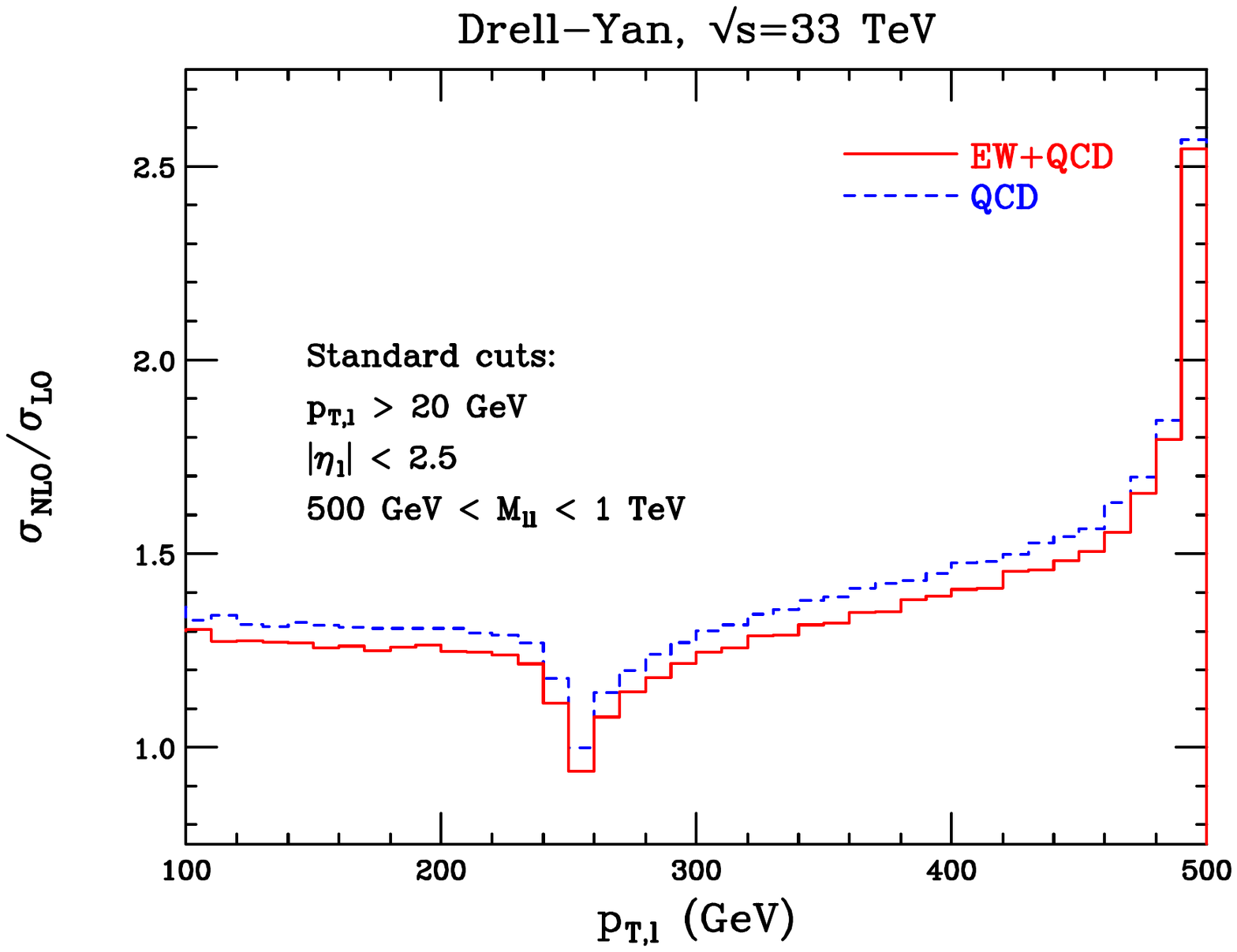}
   \includegraphics[width=0.49\textwidth]{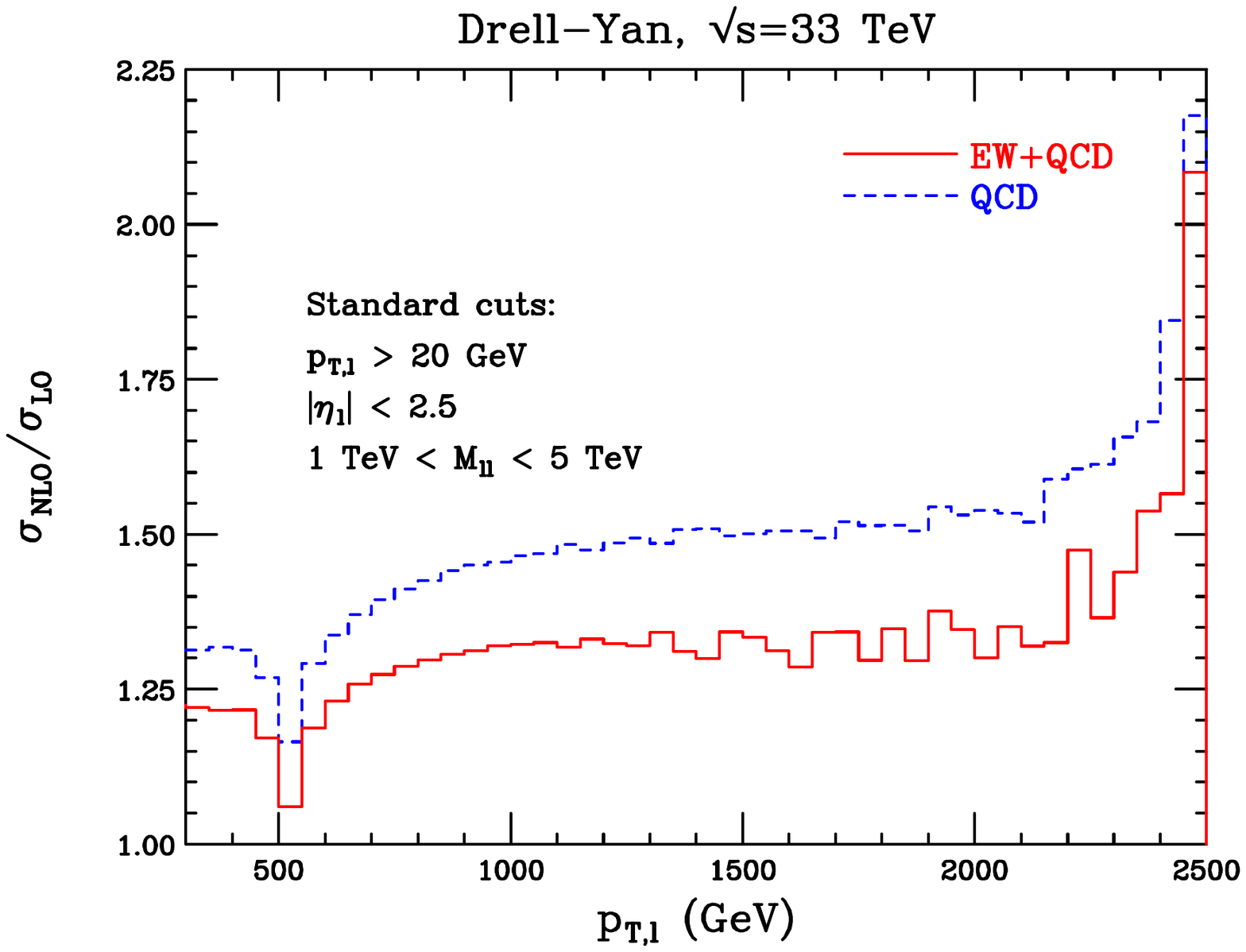}
   \vspace{-1.0in}
   \caption{Inclusive vector boson production: QCD corrections and combined electroweak-QCD corrections to lepton-pair production as a function of the lepton transverse momentum, at a 33~TeV $pp$ collider.  Results for two bins of lepton-pair invariant mass, $M_{ll} \in [500\, {\rm GeV}, 1 \, {\rm TeV}]$, and $M_{ll} \in [1\, {\rm TeV}, 5 \, {\rm TeV}]$, are shown.}
   \label{fig:qcd-EW-Flep1}
\end{figure}

\begin{figure}[h!]
   \centering
   \includegraphics[width=0.49\textwidth]{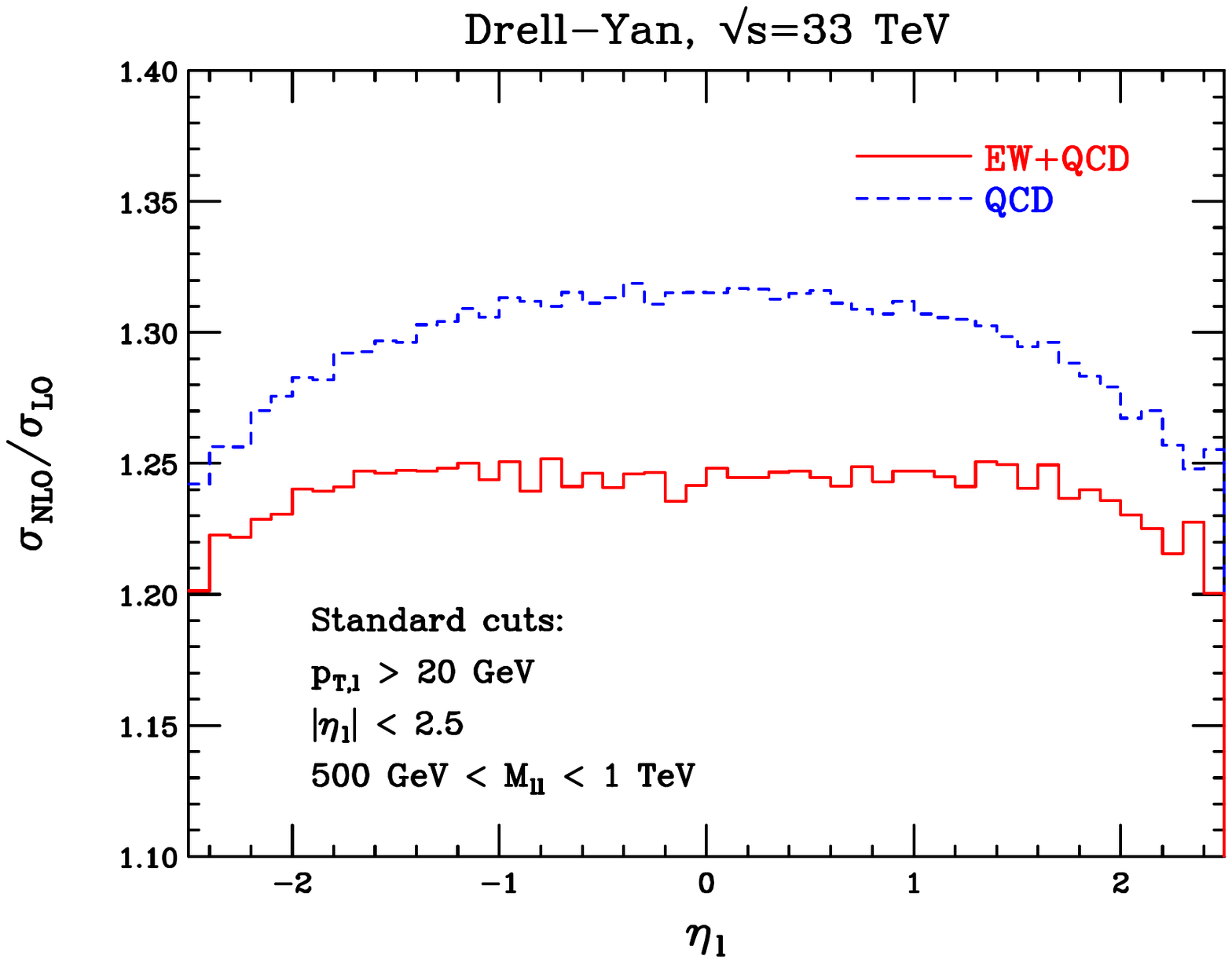}
   \includegraphics[width=0.49\textwidth]{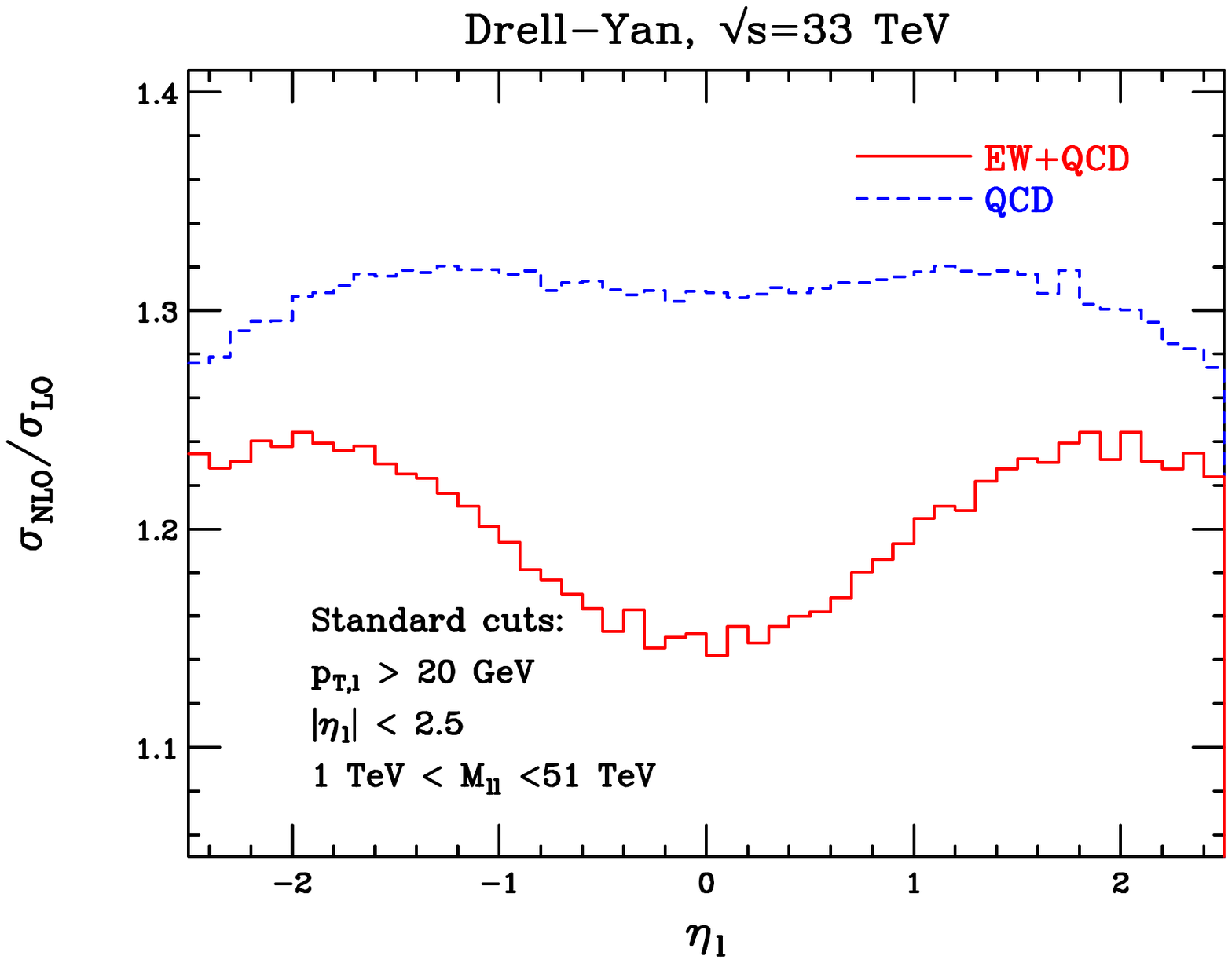}
   \vspace{-1.0in}
   \caption{Inclusive vector boson production: QCD corrections and combined electroweak-QCD corrections to lepton-pair production as a function of the lepton pseudorapidity, at a 33~TeV $pp$ collider.  Results for two bins of lepton-pair invariant mass $M_{ll} \in [500\, {\rm GeV}, 1 \, {\rm TeV}]$, and $M_{ll} \in [1\, {\rm TeV}, 5 \, {\rm TeV}]$, are shown.}
   \label{fig:qcd-EW-Flep2}
\end{figure}

\begin{figure}[h!]
   \centering
   \includegraphics[width=0.49\textwidth]{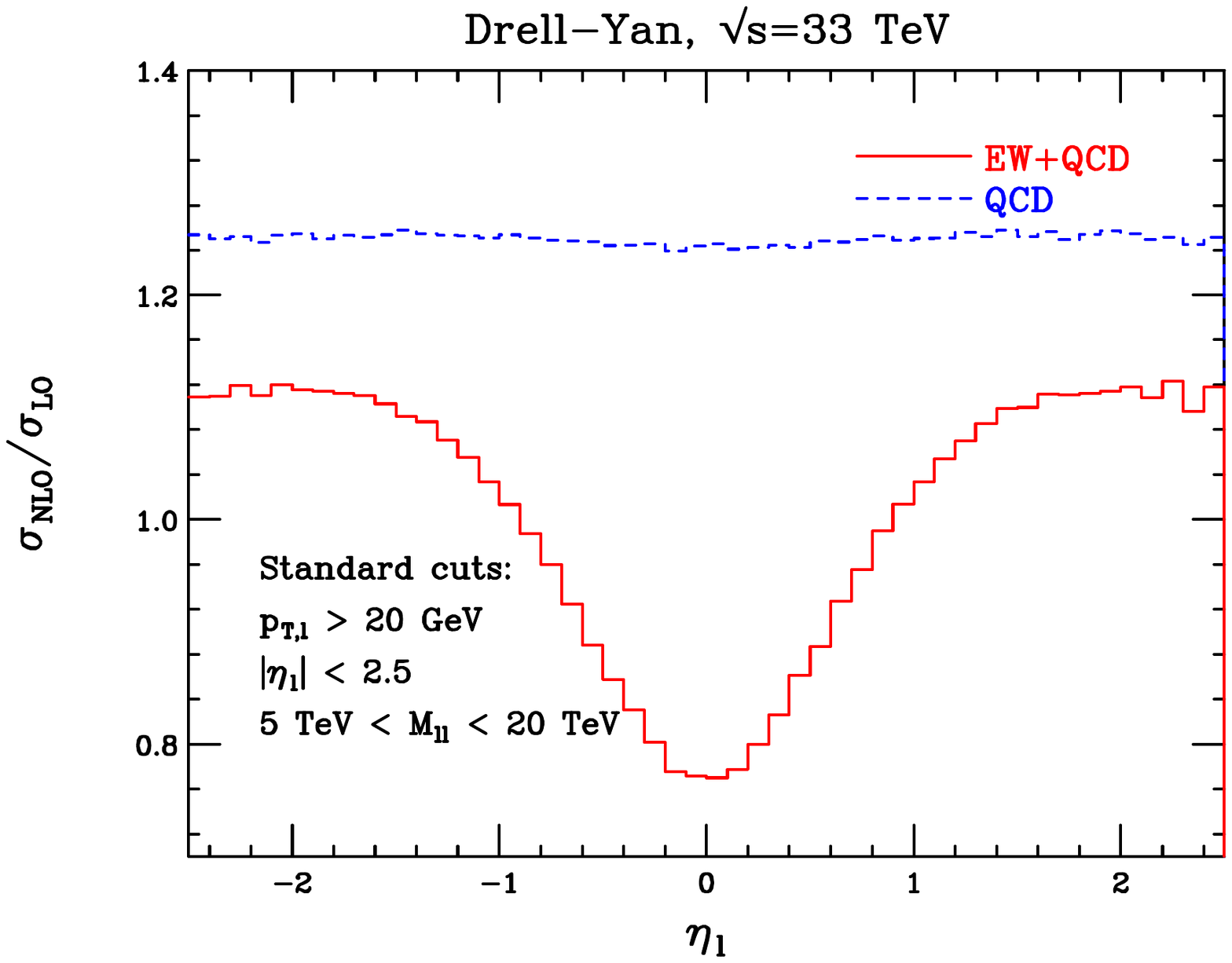}
   \includegraphics[width=0.49\textwidth]{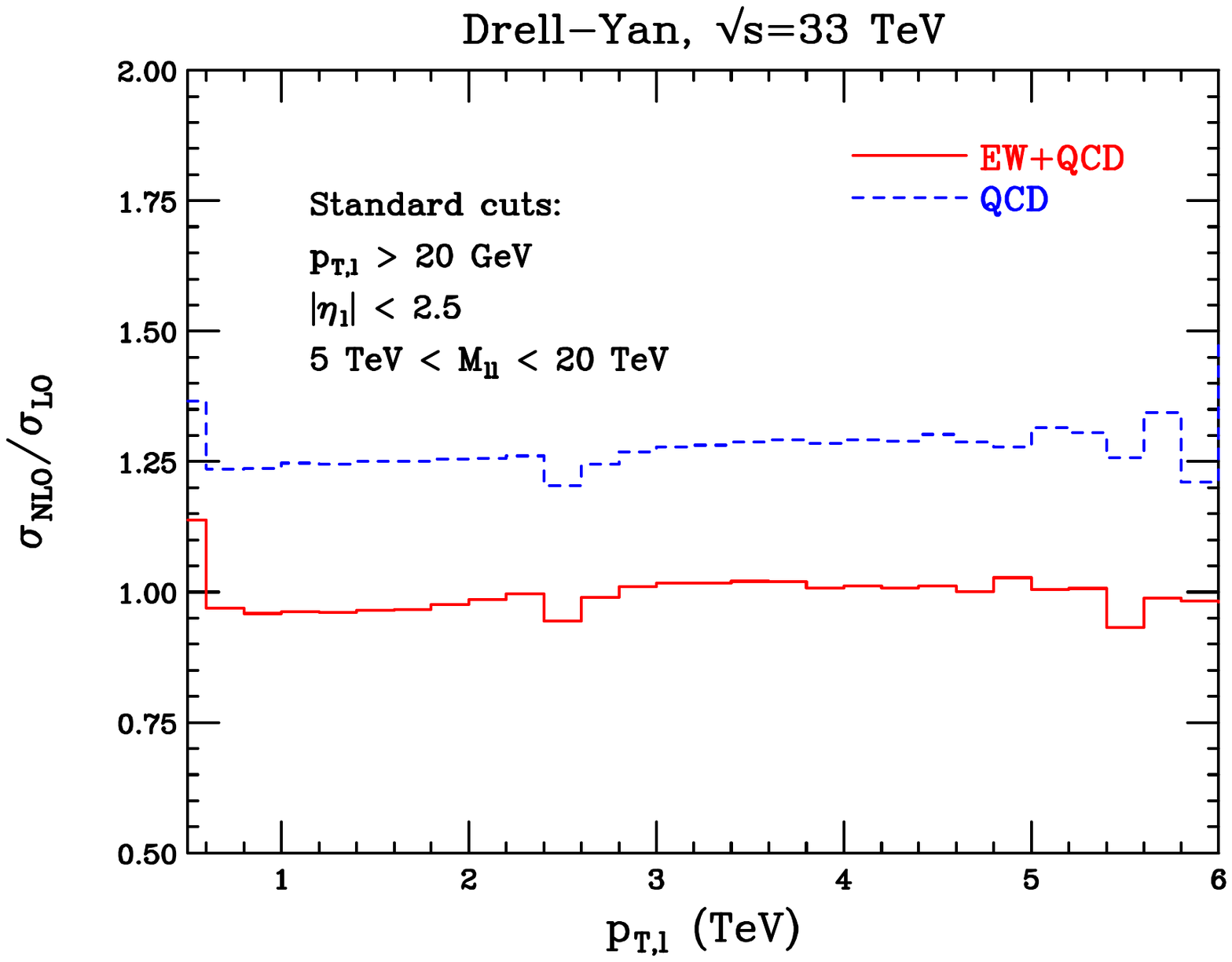}
   \vspace{-1.0in}
   \caption{Inclusive vector boson production: QCD corrections and combined electroweak-QCD corrections to lepton-pair production as a function of the lepton pseudorapidity and transverse momentum, at a 33~TeV $pp$ collider.  Results for the bin of lepton-pair invariant mass  $M_{ll} \in [5\, {\rm TeV}, 20 \, {\rm TeV}]$ are shown}
   \label{fig:qcd-EW-Flep3}
\end{figure}

The shifts induced by the combined QCD and electroweak corrections on the lepton differential distributions are considered next.  The cross section is first divided into the following invariant mass bins: $M_{ll} \in [500\, {\rm GeV}, 1 \, {\rm TeV}]$, $M_{ll} \in [1\, {\rm TeV}, 5 \, {\rm TeV}]$, and $M_{ll} \in [5\, {\rm TeV}, 20 \, {\rm TeV}]$.  The lepton transverse momentum and pseudorapidity distributions in each bin are then studied.  The results are shown in Figs.~\ref{fig:qcd-EW-Flep1},~\ref{fig:qcd-EW-Flep2},~and~\ref{fig:qcd-EW-Flep3}.  The QCD and electroweak corrections have the same shape as a function of lepton $p_T$.  The dips appearing in the corrections at half the lower bin edge, and the rise at the upper bin edge, are artifacts of the Jacobian peaks present in the leading-order result.  An interesting feature emerges in the lepton $\eta$ distributions at higher invariant masses.  The electroweak corrections act more strongly for central pseudorapidities, leading to a dip in the combined corrections that is quite large for the highest invariant mass bin.  Large negative shifts in inclusive observables such as the vector-boson transverse momentum or invariant mass are not the only effects in the Sudakov zone.  Distortions in the lepton $\eta$ also occur because of the shape differences between QCD and electroweak corrections.

\section{Vector boson production in association with jets}
\begin{figure}[!htb]
\begin{center}
\includegraphics[width=0.4\textwidth]{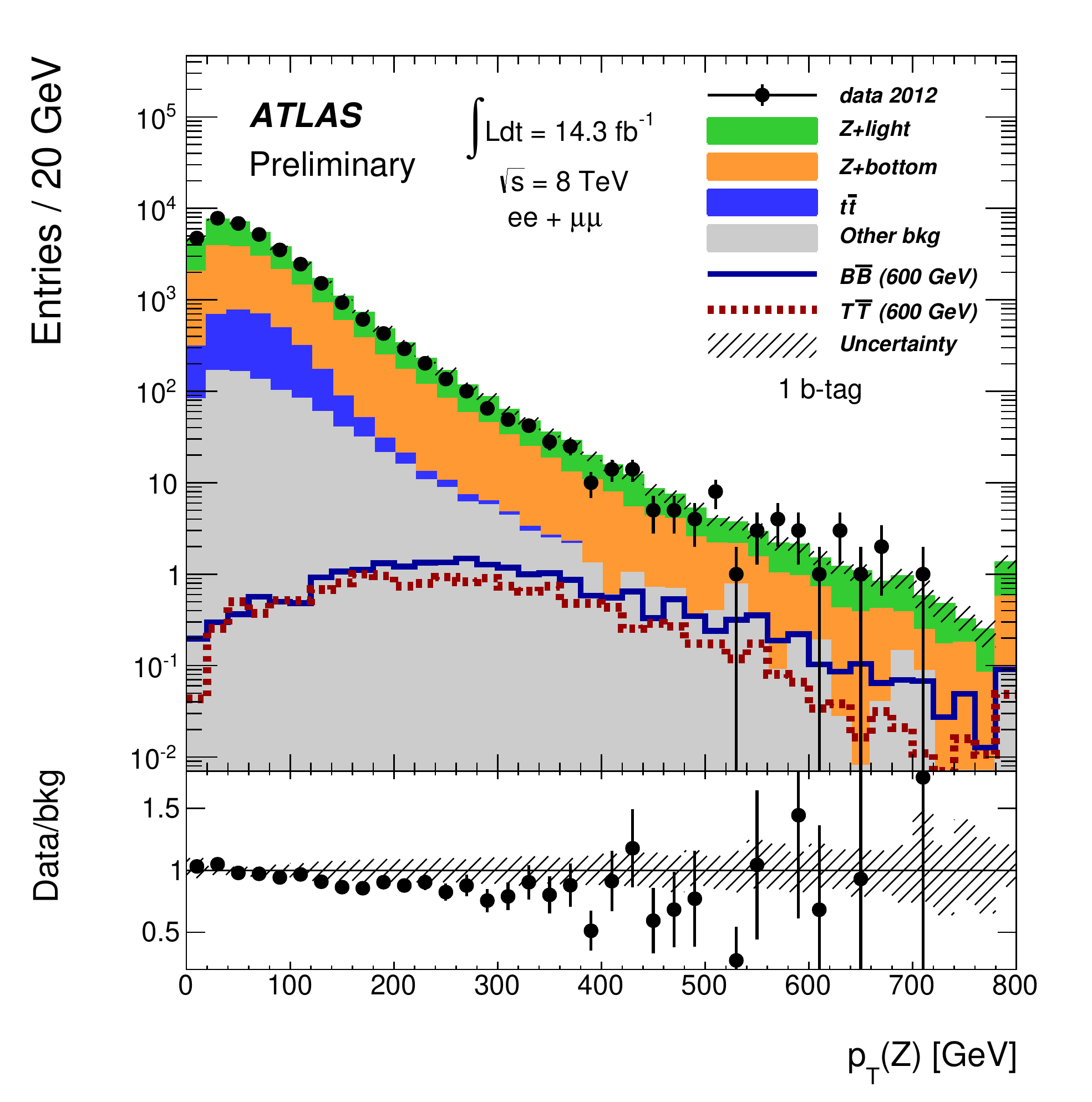}
\includegraphics[width=0.4\textwidth]{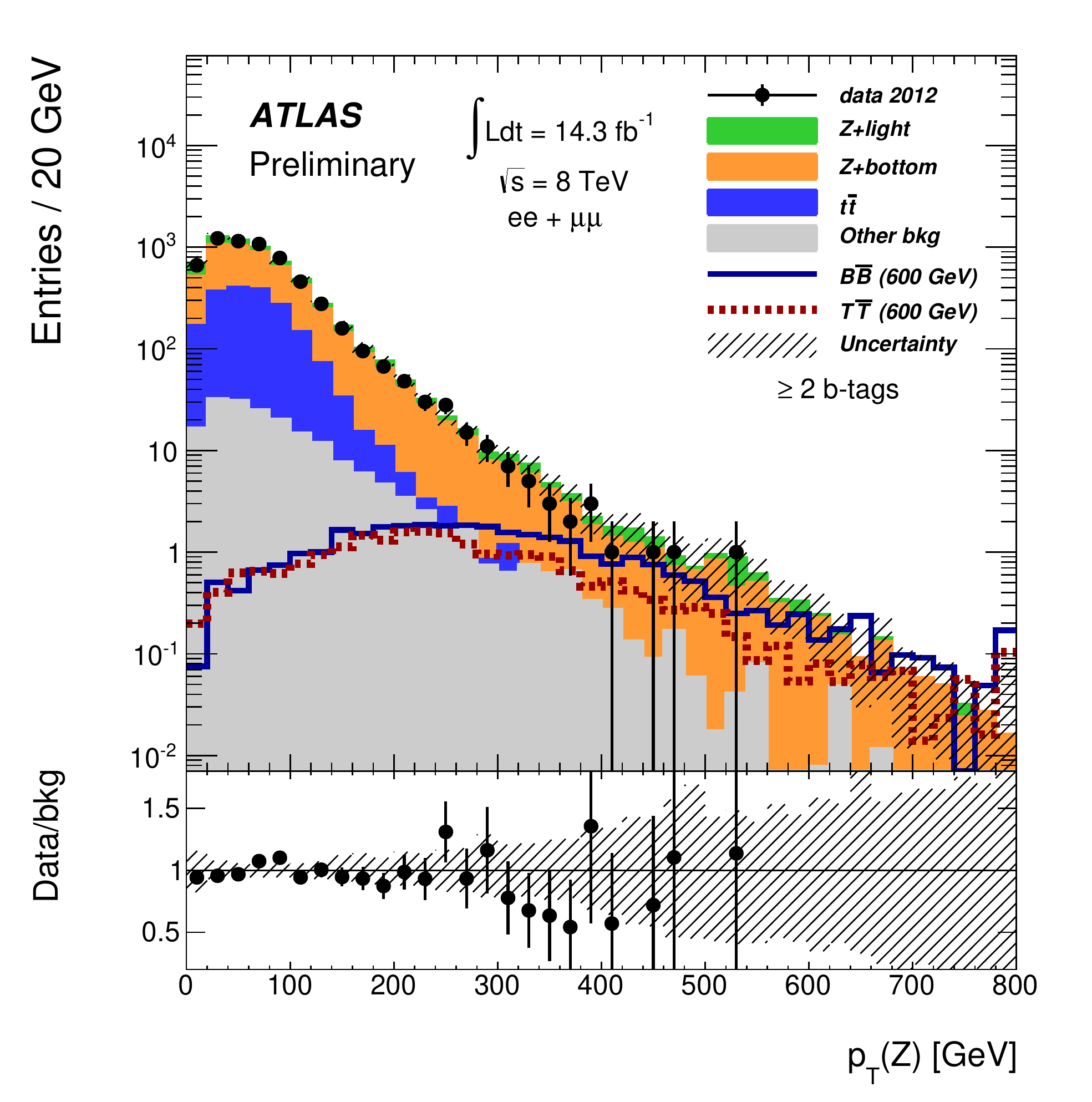}
\caption{Vector boson production in association with jets: 
  Observed Z boson transverse momentum distribution in events 
  containing at least two jets at $\sqrt{s}=8~\TeV$. 
  (left) Exactly one jet is b-tagged. 
  (right) At least two jets are b-tagged. 
  For details see Ref.~\cite{atlas:Zjets}.}
\label{fig:Vjets_exp}
\end{center}
\end{figure}
\begin{table} 
                                                                                                                                         $$ \begin{array}{c|rrrrrr}
                                                                  \multicolumn{7}{c}{\Pp\Pp
                                                                    \to l^+l^-\; \mathrm{jet} + X \;\mbox{at} \;\sqrt{s} =33 \TeV} \\
              \hline p_{\rT,\mathrm{jet}} / \GeV & 50-\infty \;\;\; & 100-\infty \;\;\; & 200-\infty \;\; & 400-\infty \;\; & 1000-\infty \;\; & 2000-\infty \; \\ 
  \hline \hline 
                 \si_{\born}^{\mu = \MZ}/\pba \; & \;              121.26(1)                  \; & \;              33.737(4)                  \; & \;              6.3314(7)                  \; & \;             0.79429(9)                  \; & \;            0.026920(2)                  \; & \;           0.0010264(1)                  \\ 
              \si_{\born}^{\mathrm{var}}/\pba \; & \;              120.70(1)                  \; & \;              32.654(4)                  \; & \;              5.6403(7)                  \; & \;             0.60157(6)                  \; & \;            0.014491(1)                  \; & \;          0.00038663(4)                  \\ 
  \hline \hline 
                   \de_{\EW}^{\mu = \MZ} / \% \; & \;               -4.68(1)                  \; & \;               -5.47(1)                  \; & \;               -9.25(1)                  \; & \;              -16.15(2)                  \; & \;              -29.25(5)                  \; & \;               -41.8(2)                  \\ 
    \de_{\EW}^{\mathrm{rec}\,,\mu = \MZ} / \% \; & \;               -3.29(3)                  \; & \;               -4.48(2)                  \; & \;               -8.50(4)                  \; & \;              -15.42(3)                  \; & \;              -28.59(8)                  \; & \;               -41.0(1)                  \\ 
 \hline 
                  \de_{\EW}^{\mathrm{var}}/\% \; & \;               -4.61(1)                  \; & \;               -5.31(1)                  \; & \;               -8.92(1)                  \; & \;              -15.65(2)                  \; & \;              -28.48(5)                  \; & \;               -40.5(3)                  \\ 
   \de_{\EW}^{\mathrm{rec}\,,\mathrm{var}}/\% \; & \;               -3.16(3)                  \; & \;               -4.42(4)                  \; & \;               -8.12(2)                  \; & \;              -14.89(2)                  \; & \;              -27.62(5)                  \; & \;               -39.9(1)                  \\ 
  \hline \hline 
                    \de_{\QCD}^{\mu = \MZ}/\% \; & \;                56.1(2)                  \; & \;                78.7(1)                  \; & \;               113.0(1)                  \; & \;               160.5(1)                  \; & \;               240.4(2)                  \; & \;               330.6(3)                  \\ 
                 \de_{\QCD}^{\mathrm{var}}/\% \; & \;                54.0(1)                  \; & \;                77.4(1)                  \; & \;               117.3(1)                  \; & \;               186.0(1)                  \; & \;               347.2(2)                  \; & \;               609.0(4)                  \\ 
 \hline 
              \de_{\QCD,\veto}^{\mu = \MZ}/\% \; & \;                13.1(1)                  \; & \;                23.3(1)                  \; & \;                29.7(1)                  \; & \;                27.5(1)                  \; & \;                -2.8(1)                  \; & \;               -42.2(1)                  \\ 
           \de_{\QCD,\veto}^{\mathrm{var}}/\% \; & \;                12.9(1)                  \; & \;                25.6(2)                  \; & \;                38.8(1)                  \; & \;                49.8(1)                  \; & \;                52.6(1)                  \; & \;                54.0(1)                  \\ 
  \hline \hline 
               \de_{\ga,\born}^{\mu = \MZ}/\% \; & \;              0.1377(4)                  \; & \;              0.1647(4)                  \; & \;              0.1960(5)                  \; & \;              0.2442(7)                  \; & \;               0.341(1)                  \; & \;               0.457(1)                  \\ 
            \de_{\ga,\born}^{\mathrm{var}}/\% \; & \;              0.1489(4)                  \; & \;              0.1923(5)                  \; & \;              0.2562(7)                  \; & \;               0.371(1)                  \; & \;               0.661(2)                  \; & \;               1.095(3)                  \\ 
 \hline\hline 
     \si_{\full,\veto}^{\mathrm{var}}/\pba/\% \; & \;               130.9(2)                  \; & \;               39.36(6)                  \; & \;               7.342(9)                  \; & \;              0.8091(7)                  \; & \;             0.01808(2)                  \; & \;            0.000443(1)                  \\ 
 \end{array} $$
  \caption{Z + 1-jet production: 
    Integrated cross sections for different cuts on the \pt 
    of the leading jet (jet with highest \pt) at a pp  
    collider with $\sqrt{s} = 33~\TeV$. 
   The LO results are shown both for a variable and for a constant scale. 
   The relative EW corrections $\de_{\EW}$ are given with and without
   lepton-photon recombination. The QCD corrections $\de_{\QCD}$ 
   are presented for a fixed as well
   as for a variable scale and with or without employing a veto on
   a second hard jet. The EW corrections and the corrections due 
   to photon-induced processes, $\de_{\ga}$, are presented for the
   variable scale. Finally, the last row shows the full NLO cross section
   $\si_{\full,\veto}^{\mathrm{var}}$. The error from the
   Monte Carlo integration for the last digit(s) is given in parenthesis. 
   See Ref.~\cite{Denner:2011vu} for details.}
\label{tab:corr:Zlljetptj33}
\end{table}
\begin{table}
                                                              $$ \begin{array}{c|rrrrr}
                                                                  \multicolumn{6}{c}{\Pp\Pp
                                                                    \to l^+l^-\; \mathrm{jet} + X \;\mbox{at} \;\sqrt{s} =33 \TeV} \\
              \hline M_{ll} / \GeV & 100-\infty \;\;\; & 200-\infty \;\;\; & 400-\infty \;\; & 1000-\infty \;\; & 2000-\infty \; \\  

 \hline \hline 
                 \si_{\born}^{\mu = \MZ}/\pba \; & \;              19.924(6)                  \; & \;              1.6890(2)                  \; & \;             0.28005(4)                  \; & \;            0.022682(3)                  \; & \;           0.0024968(4)                                    \\ 
              \si_{\born}^{\mathrm{var}}/\pba \; & \;              19.849(6)                  \; & \;              1.6482(2)                  \; & \;             0.26618(4)                  \; & \;            0.020604(3)                  \; & \;           0.0021755(3)                                   \\ 
  \hline \hline 
                   \de_{\EW}^{\mu = \MZ} / \% \; & \;                -9.6(1)                  \; & \;               -5.74(1)                  \; & \;               -8.26(1)                  \; & \;              -14.31(2)                  \; & \;              -21.69(3)                                    \\ 
    \de_{\EW}^{\mathrm{rec}\,,\mu = \MZ} / \% \; & \;                -5.3(1)                  \; & \;               -3.06(1)                  \; & \;               -5.13(1)                  \; & \;              -10.11(2)                  \; & \;              -16.34(3)                                    \\ 
 \hline 
                  \de_{\EW}^{\mathrm{var}}/\% \; & \;               -9.46(8)                  \; & \;               -5.69(1)                  \; & \;               -8.14(1)                  \; & \;              -14.18(2)                  \; & \;              -21.56(3)                                   \\ 
   \de_{\EW}^{\mathrm{rec}\,,\mathrm{var}}/\% \; & \;               -5.05(7)                  \; & \;               -2.94(1)                  \; & \;               -4.93(1)                  \; & \;               -9.93(2)                  \; & \;              -16.14(3)                                   \\ 
  \hline \hline 
                    \de_{\QCD}^{\mu = \MZ}/\% \; & \;                 29.(1)                  \; & \;                14.8(2)                  \; & \;                -0.6(1)                  \; & \;               -29.5(1)                  \; & \;               -57.1(1)                                   \\ 
                 \de_{\QCD}^{\mathrm{var}}/\% \; & \;                27.9(6)                  \; & \;                15.9(2)                  \; & \;                 2.2(1)                  \; & \;               -23.0(1)                  \; & \;               -46.6(1)                                   \\ 
 \hline 
              \de_{\QCD,\veto}^{\mu = \MZ}/\% \; & \;                 5.0(6)                  \; & \;                -8.9(2)                  \; & \;               -25.5(1)                  \; & \;               -54.9(1)                  \; & \;               -82.4(1)                                   \\ 
           \de_{\QCD,\veto}^{\mathrm{var}}/\% \; & \;                 6.1(4)                  \; & \;                -7.2(2)                  \; & \;               -20.8(2)                  \; & \;               -45.8(1)                  \; & \;               -69.8(1)                                   \\ 
  \hline \hline 
               \de_{\ga,\born}^{\mu = \MZ}/\% \; & \;               0.669(1)                  \; & \;               2.097(5)                  \; & \;               2.409(6)                  \; & \;               2.168(6)                  \; & \;               1.844(5)                                  \\ 
            \de_{\ga,\born}^{\mathrm{var}}/\% \; & \;               0.710(1)                  \; & \;               2.298(5)                  \; & \;               2.721(7)                  \; & \;               2.510(7)                  \; & \;               2.135(5)                                  \\ 
 \hline \hline
     \si_{\full,\veto}^{\mathrm{var}}/\pba/\% \; & \;               19.32(9)                  \; & \;               1.473(4)                  \; & \;              0.1965(6)                  \; & \;             0.00877(2)                  \; & \;            0.000234(3)                               \\ 
\end{array}$$

  \caption{Z + 1-jet production: 
    Integrated cross sections for different cuts on the 
    dilepton invariant mass at a pp 
    collider with $\sqrt{s} = 33~\TeV$. See aption of 
    Table~\ref{tab:corr:Zlljetptj33} and text for details.}
\label{tab:corr:ZlljetMll33}
\end{table}
\begin{table} 
                                    $$ \begin{array}{c|rrrrrr}
                                                                  \multicolumn{7}{c}{\Pp\Pp
                                                                    \to l^+l^-\; \mathrm{jet} + X \;\mbox{at} \;\sqrt{s} =100 \TeV} \\
              \hline p_{\rT,\mathrm{jet}} / \GeV & 100-\infty \;\;\; & 200-\infty \;\;\; & 400-\infty \;\; & 800-\infty \;\; & 2000-\infty \;\; & 4000-\infty \; \\ 

  \hline \hline 
                 \si_{\born}^{\mu = \MZ}/\pba \; & \;              114.29(1)                  \; & \;              23.772(3)                  \; & \;              3.5452(4)                  \; & \;             0.42003(4)                  \; & \;            0.017238(1)                  \; & \;          0.00094403(9)                  \\ 
              \si_{\born}^{\mathrm{var}}/\pba \; & \;              118.30(1)                  \; & \;              23.762(3)                  \; & \;              3.1922(3)                  \; & \;             0.31583(3)                  \; & \;           0.0091290(9)                  \; & \;          0.00035205(3)                  \\ 
  \hline \hline 
                   \de_{\EW}^{\mu = \MZ} / \% \; & \;               -5.62(1)                  \; & \;               -9.57(1)                  \; & \;              -16.86(2)                  \; & \;              -27.11(8)                  \; & \;               -43.5(1)                  \; & \;               -58.8(1)                  \\ 
    \de_{\EW}^{\mathrm{rec}\,,\mu = \MZ} / \% \; & \;               -4.65(3)                  \; & \;               -8.72(2)                  \; & \;              -16.08(2)                  \; & \;              -26.29(4)                  \; & \;              -43.15(7)                  \; & \;               -58.5(2)                  \\ 
 \hline 
                  \de_{\EW}^{\mathrm{var}}/\% \; & \;               -5.50(1)                  \; & \;               -9.29(1)                  \; & \;              -16.38(3)                  \; & \;              -26.36(4)                  \; & \;               -43.2(2)                  \; & \;               -57.5(1)                  \\ 
   \de_{\EW}^{\mathrm{rec}\,,\mathrm{var}}/\% \; & \;               -4.48(2)                  \; & \;               -8.52(2)                  \; & \;              -15.62(2)                  \; & \;              -25.64(4)                  \; & \;              -42.21(7)                  \; & \;               -56.8(1)                  \\ 
  \hline \hline 
                    \de_{\QCD}^{\mu = \MZ}/\% \; & \;                97.4(2)                  \; & \;               146.0(1)                  \; & \;               215.2(2)                  \; & \;               288.7(2)                  \; & \;               378.0(3)                  \; & \;               472.6(5)                  \\ 
                 \de_{\QCD}^{\mathrm{var}}/\% \; & \;                85.4(2)                  \; & \;               130.0(2)                  \; & \;               201.7(1)                  \; & \;               298.8(2)                  \; & \;               487.9(3)                  \; & \;               769.0(7)                  \\ 
 \hline 
              \de_{\QCD,\veto}^{\mu = \MZ}/\% \; & \;                35.7(2)                  \; & \;                54.2(1)                  \; & \;                66.7(1)                  \; & \;                61.3(1)                  \; & \;                13.2(2)                  \; & \;               -43.1(1)                  \\ 
           \de_{\QCD,\veto}^{\mathrm{var}}/\% \; & \;                29.6(2)                  \; & \;                47.4(1)                  \; & \;                65.5(1)                  \; & \;                76.4(3)                  \; & \;                65.6(2)                  \; & \;                51.5(1)                  \\ 
  \hline \hline 
               \de_{\ga,\born}^{\mu = \MZ}/\% \; & \;              0.1218(3)                  \; & \;              0.1400(4)                  \; & \;              0.1681(5)                  \; & \;              0.2114(7)                  \; & \;               0.291(1)                  \; & \;               0.382(1)                  \\ 
            \de_{\ga,\born}^{\mathrm{var}}/\% \; & \;              0.1407(3)                  \; & \;              0.1799(5)                  \; & \;              0.2482(7)                  \; & \;               0.365(1)                  \; & \;               0.630(2)                  \; & \;               1.006(5)                  \\ 
 \hline \hline 
     \si_{\full,\veto}^{\mathrm{var}}/\pba/\% \; & \;               147.0(2)                  \; & \;               32.86(4)                  \; & \;               4.767(4)                  \; & \;               0.475(1)                  \; & \;             0.01124(2)                  \; & \;           0.0003343(7)                  \\ 
\end{array} $$
  \caption{Z + 1-jet production: 
    Integrated cross sections for different cuts on the \pt 
    of the leading jet at a pp  
    collider with $\sqrt{s} = 100~\TeV$. See aption of 
    Table~\ref{tab:corr:Zlljetptj33} and text for details.}
\label{tab:corr:Zlljetptj100}
\end{table}
\begin{table}
   $$ \begin{array}{c|rrrrr} 
 \multicolumn{6}{c}{\Pp\Pp
                                                                    \to l^+l^-\; \mathrm{jet} + X \;\mbox{at} \;\sqrt{s} =100 \TeV} \\
              \hline M_{ll} / \GeV & 100-\infty \;\;\; & 200-\infty \;\;\; & 400-\infty \;\; & 1000-\infty \;\; & 2000-\infty \; \\ 

  \hline \hline 
                 \si_{\born}^{\mu = \MZ}/\pba \; & \;               60.70(2)                  \; & \;              5.3194(9)                  \; & \;              0.9142(1)                  \; & \;             0.08462(1)                  \; & \;            0.012762(2)                  \;                  \\ 
              \si_{\born}^{\mathrm{var}}/\pba \; & \;               61.76(2)                  \; & \;              5.3548(9)                  \; & \;              0.9016(1)                  \; & \;             0.07980(1)                  \; & \;            0.011500(2)                  \;                  \\ 
  \hline \hline 
                   \de_{\EW}^{\mu = \MZ} / \% \; & \;                -9.4(1)                  \; & \;               -5.85(1)                  \; & \;               -8.41(1)                  \; & \;              -14.28(2)                  \; & \;              -21.05(3)                                   \\ 
    \de_{\EW}^{\mathrm{rec}\,,\mu = \MZ} / \% \; & \;                -5.0(1)                  \; & \;               -3.23(1)                  \; & \;               -5.29(1)                  \; & \;              -10.42(3)                  \; & \;              -16.42(3)                                    \\ 
 \hline 
                  \de_{\EW}^{\mathrm{var}}/\% \; & \;                -9.5(2)                  \; & \;               -5.76(1)                  \; & \;               -8.31(1)                  \; & \;              -14.08(2)                  \; & \;              -20.85(3)                                    \\ 
   \de_{\EW}^{\mathrm{rec}\,,\mathrm{var}}/\% \; & \;                -5.2(2)                  \; & \;               -3.07(1)                  \; & \;               -5.15(2)                  \; & \;              -10.11(2)                  \; & \;              -16.22(4)                                    \\ 
  \hline \hline 
                    \de_{\QCD}^{\mu = \MZ}/\% \; & \;                 36.(4)                  \; & \;                20.8(3)                  \; & \;                 7.3(2)                  \; & \;               -18.1(1)                  \; & \;               -43.3(1)                                   \\ 
                 \de_{\QCD}^{\mathrm{var}}/\% \; & \;                29.3(9)                  \; & \;                18.4(3)                  \; & \;                 6.0(1)                  \; & \;               -16.3(1)                  \; & \;               -37.4(1)                                   \\ 
 \hline 
              \de_{\QCD,\veto}^{\mu = \MZ}/\% \; & \;                 5.0(9)                  \; & \;                -7.0(2)                  \; & \;               -21.2(2)                  \; & \;               -46.8(1)                  \; & \;               -71.8(1)                                   \\ 
           \de_{\QCD,\veto}^{\mathrm{var}}/\% \; & \;                 3.8(7)                  \; & \;                -7.4(3)                  \; & \;               -19.6(2)                  \; & \;               -41.7(1)                  \; & \;               -63.1(1)                                 \\ 
  \hline \hline 
               \de_{\ga,\born}^{\mu = \MZ}/\% \; & \;               0.532(1)                  \; & \;               1.679(4)                  \; & \;               2.019(6)                  \; & \;               1.988(6)                  \; & \;               1.792(6)                                  \\ 
            \de_{\ga,\born}^{\mathrm{var}}/\% \; & \;               0.568(1)                  \; & \;               1.871(4)                  \; & \;               2.363(7)                  \; & \;               2.464(7)                  \; & \;               2.259(7)                                   \\ 
 \hline \hline
     \si_{\full,\veto}^{\mathrm{var}}/\pba/\% \; & \;                58.6(4)                  \; & \;                4.75(1)                  \; & \;               0.672(1)                  \; & \;              0.0373(1)                  \; & \;             0.00211(2)                                 \\ 

\end{array}$$

  \caption{Z + 1-jet production: 
    Integrated cross sections for different cuts on the 
    dilepton invariant mass at a pp  
    collider with $\sqrt{s} = 100~\TeV$. See aption of 
    Table~\ref{tab:corr:Zlljetptj33} and text for details.}
\label{tab:corr:ZlljetMll100}
\end{table}
Production of a W or Z boson in association with jets 
has played a special role in collider physics. It was the 
dominant background to top-quark pair production at the Tevatron. 
At the LHC, it remains an important background for processes 
containing a lepton, missing energy (\met), and one or more 
jets in the final state. Prominent examples are 
measurements of top quark, Higgs boson, and multi-boson production 
and searches for supersymmetry signatures. 
Such measurements also permit stringent tests of the predictions
of the SM.
Measurements of W and Z 
boson production in association with multiple jets have been made by
the ATLAS~\cite{Aad:2012en,Aad:2011qv,atlas:Zjets} and 
CMS~\cite{Chatrchyan:2011ne,Chatrchyan:2012vr,cms:Wbb} 
collaborations at $\sqrt{s} = 7$ and $8~\TeV$. 
These  measurements show 
sensitivity to boson and leading-jet \pt of up to about $500~\GeV$ at the LHC. 
As shown in Fig.~\ref{fig:Vjets_exp}, at the 
current level of experimental and theoretical accuracy,
the SM is able to describe data well. 
The current experimental uncertainty for the highest \pt bins 
is somewhat larger than the size of the EW corrections.  
The same measurements at 14~TeV will be sensitive to 
probing the Sudakov zone.

Production cross sections for W + up to 5 jets and Z + up to 3 jets  are 
now known at NLO QCD~\cite{Berger:2010zx,Bern:2013gka}. 
The full NLO EW corrections for W + 1-jet production have been
computed for the final state containing a charged
lepton, a neutrino, and a hard jet~\cite{Denner:2009gj}, 
and for the monojet scenarios where the Z boson decays into two undetected 
neutrinos~\cite{Denner:2012ts}.
The full NLO EW corrections for Z + 1-jet production have also been
computed for the final state containing two charged
leptons and a hard jet~\cite{Denner:2011vu}.
The overall magnitude of these corrections as a function 
of the boson \pt is similar to the inclusive W/Z case.
Results from repeating the same calculation as in 
Ref.~\cite{Denner:2011vu} for Z + 1-jet process at  
$\sqrt{s} = 33~\TeV$ are listed in Tables~\ref{tab:corr:Zlljetptj33} 
and \ref{tab:corr:ZlljetMll33} as a function of 
the leading jet \pt  and dilepton invariant mass, respectively.
Corresponding results for $\sqrt{s} = 100~\TeV$ are listed in 
Tables~\ref{tab:corr:Zlljetptj100} and \ref{tab:corr:ZlljetMll100}.
We find that the relative corrections show very weak dependence   
on $\sqrt{s}$ and depend mostly on 
the jet (or boson) \pt and the invariant mass of the boson system. 
However, as the kinematic reach increases  
with $\sqrt{s}$, these corrections will become more important.  

\subsection{$Z$ production in association with two and three jets}
The virtual NLO EW Sudakov logarithmic corrections to vector bosons plus jets 
have been implemented in {\tt ALPGEN}~\cite{Chiesa:2013yma} through the 
algorithm of Refs.~\cite{Denner:2000jv,Denner:2001gw}, with phenomenological results 
for $Z + 2$ and $Z + 3$~jets production at LHC (at $7$ and $14$~TeV). 
Here we present the scaling of the corrections with the center of mass (c.m.) energy 
of proton-proton collisions from $14$~TeV 
to $33$ and $100$~TeV. The event selections, 
parameters and considered observables are those of Ref.~\cite{Chiesa:2013yma}. The 
parameters and PDF setting are the  {\tt ALPGEN} defaults. 
In particular, for $Z + 2$~jets, we consider the observables/cuts presently 
adopted by ATLAS~\cite{Aad:2011ib}, namely 
\begin{eqnarray}
&& m_{\rm eff} > 1~{\rm TeV} \qquad \, \, \, \, \, 
\rlap\slash{\!E_T}/m_{\rm eff} > 0.3  \nonumber\\
&& p_T^{j_1} > 130~{\rm GeV} \qquad \, \, 
p_T^{j_2} > 40~{\rm GeV} \quad \, \, |\eta_{j}| < 2.8 \nonumber \\
&& \Delta\phi ({\vec p}_T^j,\rlap\slash{\!\vec{p}_T}) > 0.4 \quad 
\Delta R_{(j_1, j_2)} > 0.4  \, 
\label{eq:atlascut}
\end{eqnarray}
where $j_1$ and $j_2$ are the leading and next-to-leading $p_T$ jets; 
$m_{\rm eff} = \sum_i  {p_T}_i   + \rlap\slash{\!E_T}$. 

For the $Z+3$~jets final state we consider the observables/cuts used by 
CMS~\cite{Collaboration:2011ida,Chatrchyan:2012lia}, namely
\begin{eqnarray}
&& H_T > 500~{\rm GeV} \qquad \, \, \, \, \, \, 
|\rlap\slash{\!\vec{H}_T}| > 200~{\rm GeV}  \nonumber\\
&& p_T^j > 50~{\rm GeV} \qquad  \, \, |\eta_j| < 2.5 
\quad \Delta R_{(j_i, j_k)} > 0.5 \nonumber\\
&& \Delta\phi ({\vec p}_T^{j_1, j_2},\rlap\slash{\!{\vec H}_T}) > 0.5 
\qquad  \Delta\phi (\vec{p}_T^{j_3},\rlap\slash{\!{\vec H}_T}) > 0.3 \, ,
\label{eq:cmscut}
\end{eqnarray}
where $H_T = \sum_i {p_T}_i$ and 
$\vec{\rlap\slash{\!H_T}} = 
- \sum_i \vec{p_t}_i$. 
For the sake of reference, we report here some partial results 
from Ref.~\cite{Chiesa:2013yma} corresponding to the c.m. energy of $14$~TeV. 
\begin{figure}[h]
\includegraphics[scale=0.6]{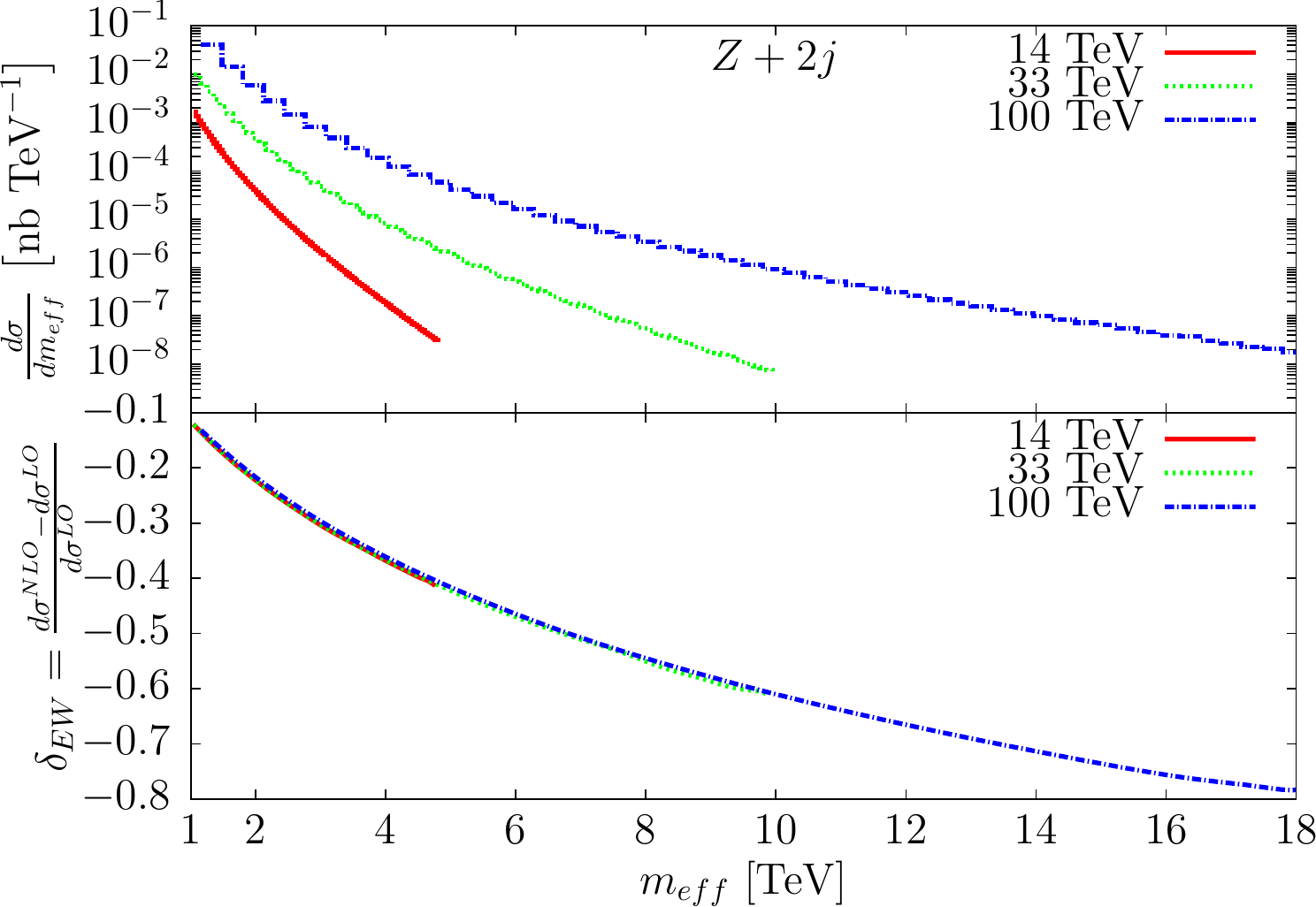}
\caption{\label{fig:meff2j} $Z+2$~jets: $d\sigma/dm_{\rm eff}$ 
(absolute LO predictions and relative EW Sudakov corrections) 
at $\sqrt{s} = 14$, $33$ and $100$~TeV.}
\end{figure}

In Fig.~\ref{fig:meff2j} 
we show the absolute LO distribution $d\sigma/dm_{\rm eff}$ and the relative 
effects of the Sudakov EW corrections corresponding to three different c.m. energies: 
$14$, $33$ and $100$~TeV. 
As can be seen, the negative corrections due to Sudakov logs are of the order of 
some tens of per cent, raising to about 40\% (60\%, 80\%) in the extreme regions at 
$\sqrt{s} = 14$ ($33$, $100$)~TeV, respectively. 
For a given bin of the $m_{\rm eff}$ distribution, the relative EW corrections 
are practically the same, independently of the collider c.m. energy.

\begin{figure}[h]
\includegraphics[scale=0.6]{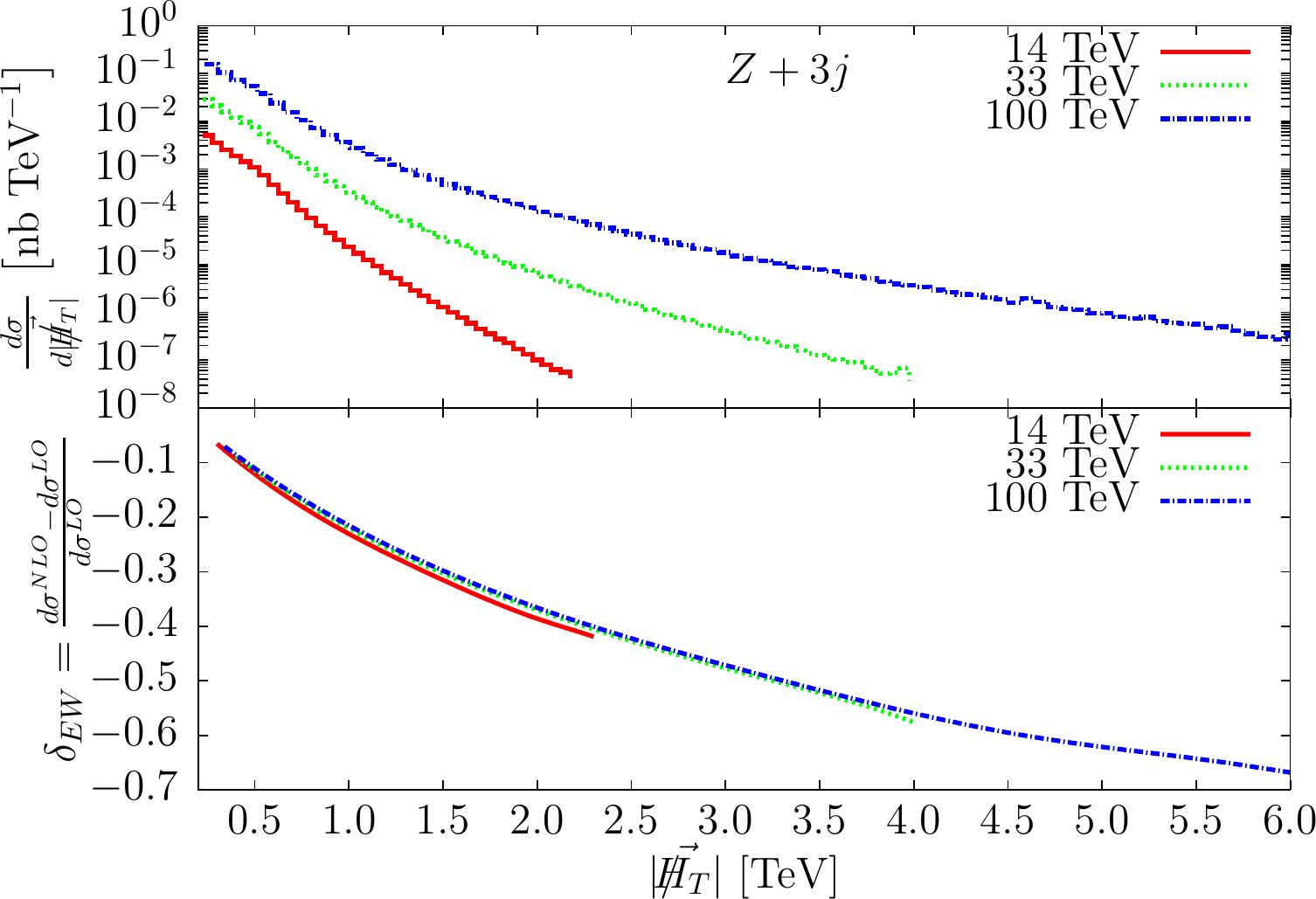}
\caption{\label{fig:Z3j} $Z+3$~jets: $d\sigma/|\rlap\slash{\!H_T}|$ 
(absolute LO predictions and relative EW Sudakov corrections) 
at $\sqrt{s} = 14$, $33$ and $100$~TeV.}
\end{figure}

In Fig.~\ref{fig:Z3j} 
we show the effect of the Sudakov logs on the observable $|\rlap\slash{\!{\vec H}_T}|$  
in the process $p p \to Z+3$~jets according to the event selection of Eq.~(\ref{eq:cmscut}). 
As for the $Z + 2$~jets effective mass distributions, the effect of NLO weak corrections 
on $|\rlap\slash{\!{\vec H}_T}|$ is large and negative, 
raising to about 40\% (60\%, 70\%) in the extreme regions at 
$\sqrt{s} = 14$ ($33$, $100$)~TeV, respectively. For a chosen $|\rlap\slash{\!{\vec H}_T}|$ bin, 
the relative effects of the corrections are quite insensitive to the change of the 
collider energy.

\section{Vector-boson pair production}
\begin{figure}[!htb]
\begin{center}
\includegraphics[width=0.45\textwidth]{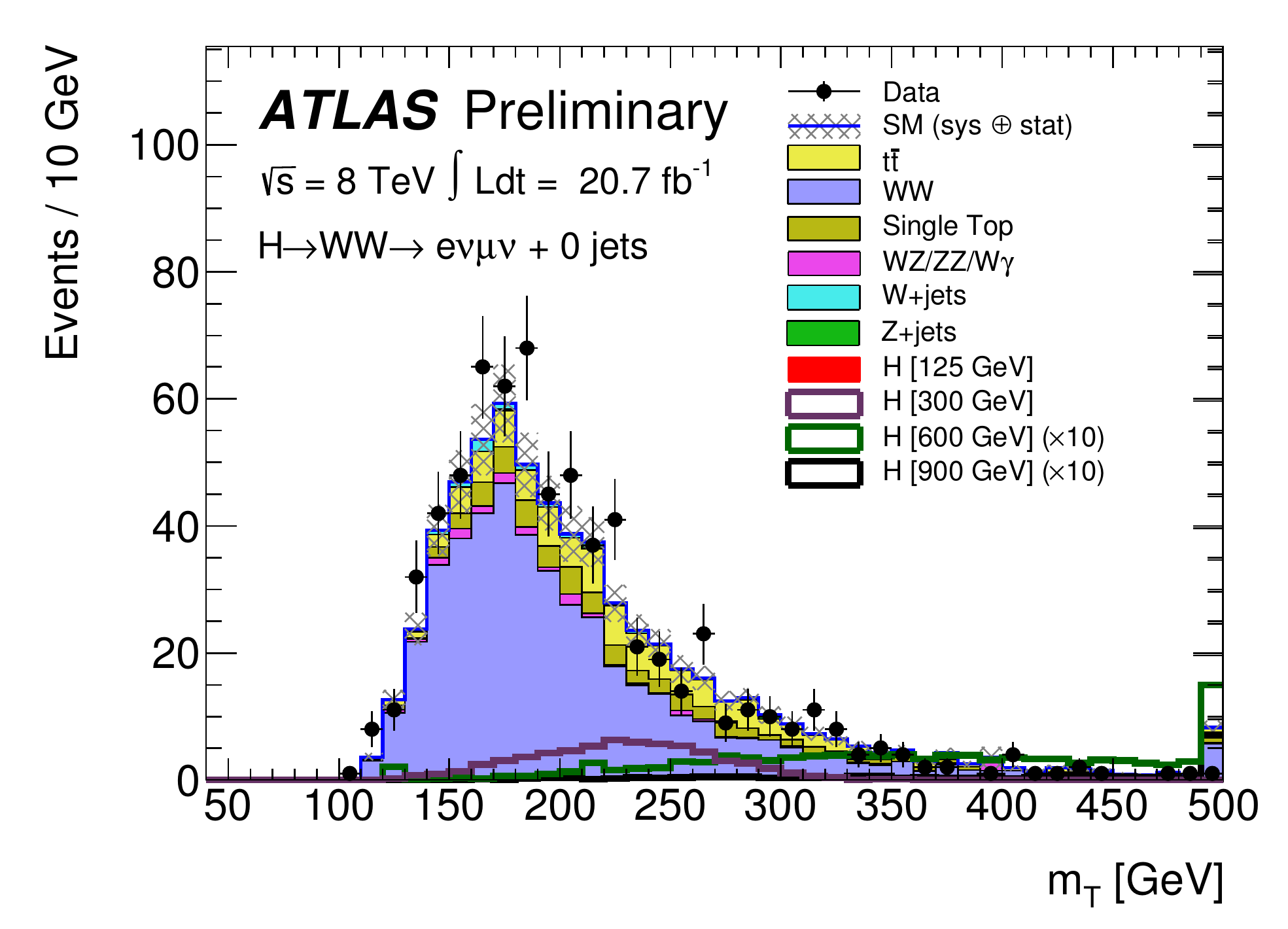}
\includegraphics[width=0.45\textwidth]{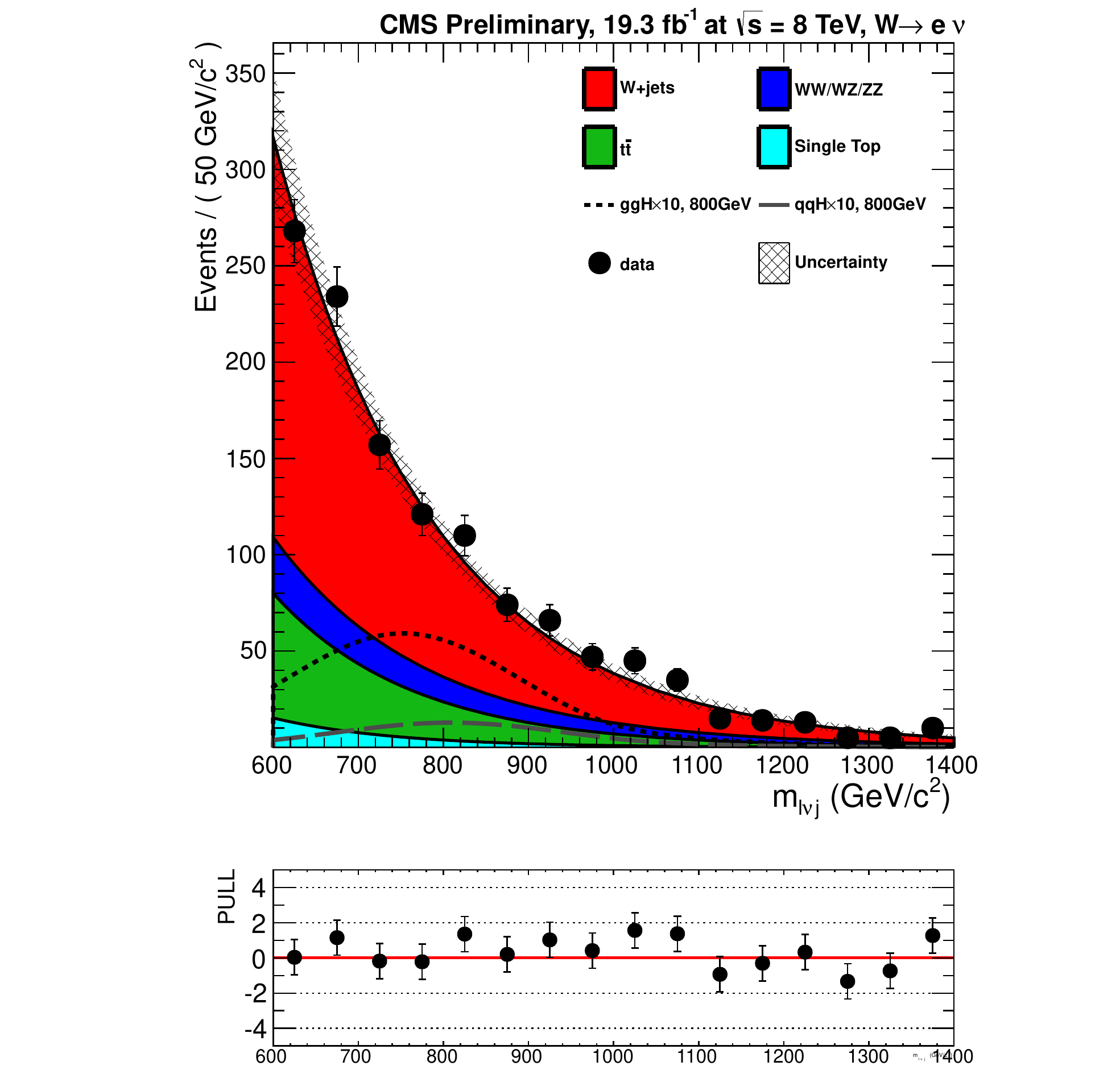}
\caption{Vector-boson pair production at $\sqrt{s}=8~\TeV$: 
  (left) Observed transverse mass distribution of the WW system 
  in the fully leptonic decay mode electron+muon+\met and no 
  jet activity. For details see Ref.~\cite{atlas:WWllvv}.
  (b) Observed invariant mass distribution of the WW candidates 
  in the semi-leptonic decay mode where one W boson decays leptonically 
  to $e\nu$ and the other W boson decays at high \pt 
  to $q\bar{q}$ giving rise to a single merged jet. 
  For details see Ref.~\cite{cms:WWlnuqq}.
}
\label{fig:VV_exp}
\end{center}
\end{figure}
\begin{figure}[!htb]
\begin{center}
\includegraphics[width=0.8\textwidth]{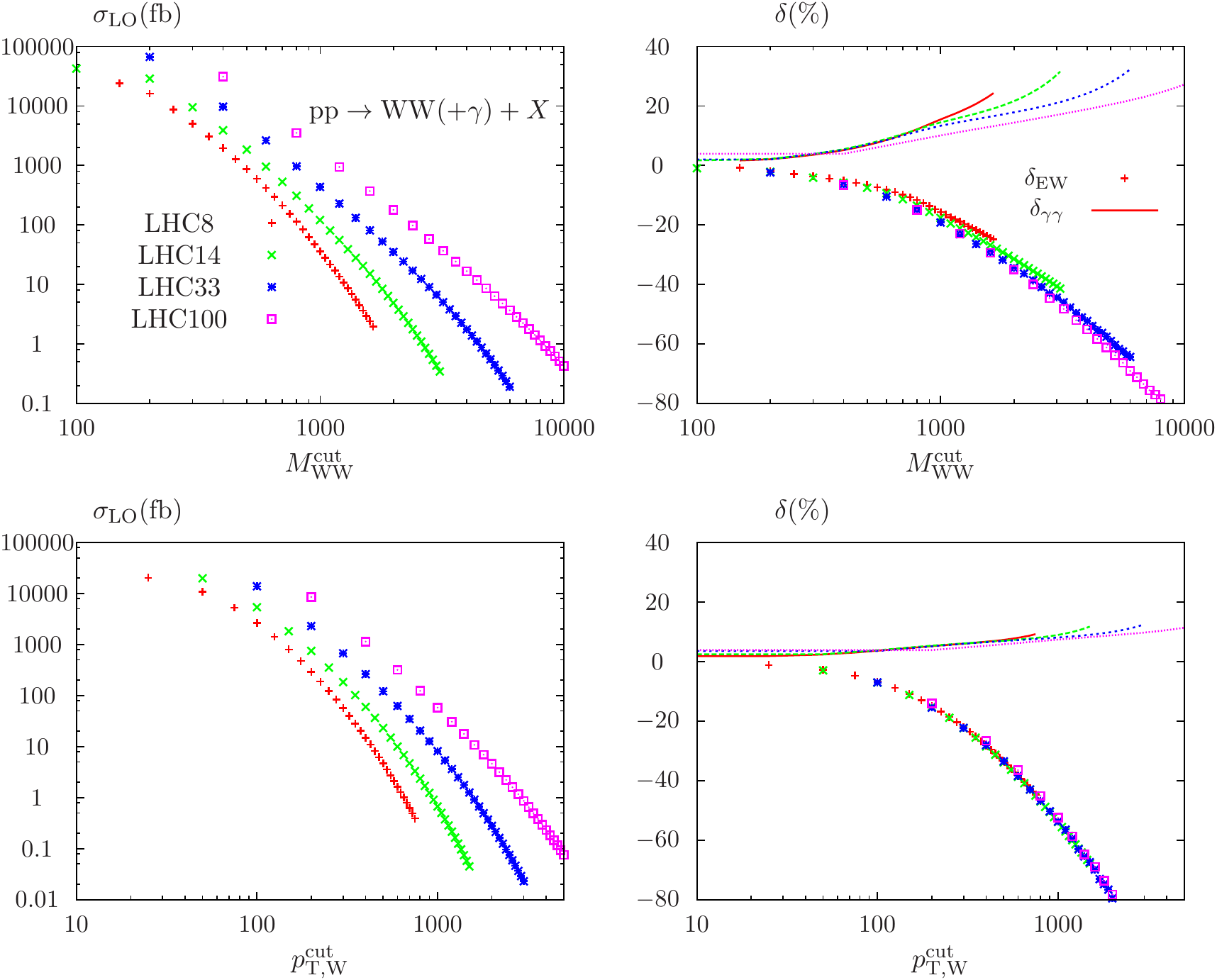}
\caption{WW production: Total cross sections for W-pair 
  production for different cuts on WW invariant mass  (top row) and 
  W transverse momentum (bottom row), evaluated at pp collision energies 
  of 8, 14, 33, and 100~TeV. 
    Left: absolute predictions, 
    right: relative electroweak corrections ($\delta_{\mathrm{EW}}$) 
    and relative contributions from the $\gamma\gamma$-induced process 
    ($\delta_{\gamma\gamma}$).}
\label{fig:WW_corr_33tev}
\end{center}
\end{figure}
Vector-boson pair production is among the most important 
SM benchmark processes at the LHC because of its 
connection to electroweak symmetry breaking, 
to the study of Higgs boson in 
H~$\to$~WW${}^*$, ZZ${}^*$, $\gamma\gamma$ processes, 
and to probes of gauge boson self interactions.  
Diboson production can also provide a deeper 
understanding of the electroweak interactions in 
general, and to test the validity of the SM at 
the highest available energies.

In this report we focus on WW production at 
large invariant masses or large W \pt, which is the 
kinematic regime of high interest that can have 
large EW corrections. 
This kinematic regime has recently been analyzed by the
ATLAS~\cite{atlas:WWllvv} and CMS~\cite{cms:WWlnuqq} 
collaborations at $\sqrt{s} = 8~\TeV$ showing 
sensitivity to invariant masses of up
to $1~\TeV$ and boson transverse momenta of up to $500~\GeV$.
Comparing with the full one-loop EW corrections 
to on-shell WW production~\cite{Bierweiler:2012kw} we find that 
the size of the correction is comparable 
to the experimental uncertainty for highest kinematic end points. 
As noted in Ref.~\cite{Bierweiler:2012kw}, the corrections 
due to photon-induced channels can be large at high enrgies, 
while radiation of additional massive vector bosons does not 
influence the results significantly. 
Recently, a detailed study of vector-boson pair production at 
NLO has been presented in Ref.~\cite{Baglio:2013toa}, where the 
authors emphasize the importance of photon-induced processes, 
which may potentially lead to significant distortions of 
differential distributions.
Results from repeating the calculation from Ref.~\cite{Bierweiler:2012kw} for  
$\sqrt{s} = 33~\TeV$ and 100~TeV are shown in Fig.~\ref{fig:WW_corr_33tev}.
We find that 
while the relative NLO EW corrections hardly depend on the 
collider energy, the relative photon-induced contributions 
are suppressed at larger values of $\sqrt{s}$. 
As a result, the overall corrections show very little dependence   
on $\sqrt{s}$. 
However, like in the case of other processes described earlier, 
the EW corrections will progressively become more important with   
increase in $\sqrt{s}$ due to higher kinematic reach.

\section{Summary}
We have presented a survey of the most abundant processes at the LHC 
for sensitivity to electroweak corrections at various pp  
collision energies relevant for LHC and future hadron colliders. 
We summarize our observations in Table~\ref{tab:summary}. 
We find that for most processes the overall electroweak 
corrections do not change much with increase in collider energy.  
However, the corrections become more important at high collider 
energies simply because of the increase in kinematic reach at 
high $\sqrt{s}$, where the corrections are inherently large. 
\begin{table}[htbp]
\begin{center}
\caption{Are we in the Sudakov zone yet?}
\label{tab:summary}
\begin{tabular}{l c  c  c} \hline \hline
Process & $\sqrt{s}=8~\TeV$ &  $\sqrt{s}=14~\TeV$  & $\sqrt{s}=33, 100~\TeV$ \\
\hline
Inclusive jet, dijet &  Yes  & Yes  & Yes  \\  \hline
Inclusive W/Z tail   &  $\sim$ Yes  & Yes  & Yes \\
W$\gamma$, Z$\gamma$ tail ($\ell\nu\gamma, \ell\ell\gamma$) & No &  $\sim$ Yes &  Yes \\  
W/Z+jets tail & $\sim$ Yes & Yes  & Yes  \\  \hline
WW leptonic  & Close & $\sim$ Yes  & Yes  \\  
WZ, ZZ leptonic  & No & No  & Yes  \\ 
WW, WZ, ZZ semileptonic & $\sim$ Yes & Yes  & Yes  \\  \hline
\end{tabular}
\end{center}
\end{table}

\section{Acknowledgements}
We would like to thank John Campbell, Kenichi Hatakeyama,  
Joey Huston, Michael Peskin, and Doreen Wackeroth for 
requesting these studies for Snowmass 2013 report and for 
providing valuable feedback.
Fermilab is operated by Fermi Research Alliance, LLC, 
under Contract No. DE-AC02-07CH11359 with the United 
States Department of Energy.
This work is supported in part by the Research Executive Agency (REA) of the 
European Union under the Grant Agreement number 
PITN-GA-2010-264564 (LHCPhenoNet), and by the Italian Ministry of University and 
Research under the 2010-2011 PRIN program 2010YJ2NYW. 
The work of L.~B. is supported by the ERC grant 291377, ``LHCtheory - 
Theoretical predictions and analyses of LHC physics: advancing the precision frontier".
The work of T.~B. and X.~G. is supported by the Swiss National 
Science Foundation (SNF) under grant  200020-140978 and the 
Sinergia grant number CRSII2 141847 1.
The work of F.~Petriello is supported by the U.S. Department of Energy, 
Division of High Energy Physics, under contract DE-AC02-06CH11357 and the 
grant DE-SC0010143.

\end{document}